\documentclass[3p,twocolumn]{elsarticle}

\journal{ArXiv}

\usepackage[utf8]{inputenc} 
\usepackage[T1]{fontenc}    
\usepackage{hyperref}       
\usepackage{url}            
\usepackage{booktabs}       
\usepackage{amsfonts}       
\usepackage{nicefrac}       
\usepackage{microtype}      
\usepackage{lipsum}
\usepackage{wrapfig}
\usepackage{multirow}

\usepackage{amssymb}
\usepackage{caption}
\usepackage{subcaption}

\usepackage{xcolor}
\usepackage{siunitx}
\usepackage{amsmath}

\DeclareMathOperator*{\minimize}{minimize}

\newcommand{\sect}[1]{Sec.~\ref{#1}}
\newcommand{\fig}[1]{Fig.~\ref{#1}}
\newcommand{\eq}[1]{Eq.~\eqref{#1}}

\newcommand{\tab}[1]{Tab.~\ref{#1}}

\usepackage{tikz}

\newcommand{\cblock}[3]{
  \hspace{-1.5mm}
  \begin{tikzpicture}
    [
    node/.style={square, minimum size=10mm, thick, line width=0pt},
    ]
    \node[fill={rgb,255:red,#1;green,#2;blue,#3}] () [] {};
  \end{tikzpicture}%
}

\newcommand{\cone}[0]{\alpha}
\newcommand{\clock}[0]{\delta}

\newcommand{\roll}[0]{\phi}
\newcommand{\sailmass}[0]{m_\text{ss}}
\newcommand{\sailradius}[0]{r_\text{ss}}
\newcommand{\sailarea}[0]{A_\text{ss}}
\newcommand{\rollsailmass}[0]{m^\phi_\text{ss}}
\newcommand{\rollsailradius}[0]{r^\phi_\text{ss}}
\newcommand{\payloadmass}[0]{m_\text{p}}

\newcommand{\xpayload}[0]{x_\text{p}}
\newcommand{\ypayload}[0]{y_\text{p}}
\newcommand{\zpayload}[0]{z_\text{p}}
\newcommand{\xsail}[0]{x_\text{ss}}
\newcommand{\ysail}[0]{y_\text{ss}}
\newcommand{\zsail}[0]{z_\text{ss}}
\newcommand{\xrollsail}[0]{x_\text{ss}^\phi}
\newcommand{\yrollsail}[0]{y_\text{ss}^\phi}

\begin{document}


\begin{frontmatter}

\address[label1]{Berkeley Sensors and Actuator Center, Berkeley, CA 94720, USA.}
\address[label4]{Mechanical Engineering Department, University of California, Berkeley, Berkeley, CA 94720, USA. }
\address[label2]{Electrical Engineering and Computer Sciences Department, University of California, Berkeley, Berkeley, CA 94720, USA. }
\address[label5]{Space Sciences Laboratory, 7 Gauss Way, Berkeley, CA 94720, USA.}

\title{BLISS: Interplanetary Exploration with Swarms of Low-Cost Spacecraft}
\author{Alexander N. Alvara \fnref{label1,label4} }
\author{Lydia Lee \fnref{label1,label2}}
\author{Emmanuel Sin \fnref{label4}}
\author{Nathan Lambert \fnref{label1,label2}}
\author{Andrew J. Westphal \fnref{label5}}
\author{Kristofer S.J. Pister \fnref{label1,label2}}



\begin{abstract}
Leveraging advancements in micro-scale technology, we propose a fleet of autonomous, low-cost, small solar sails for interplanetary exploration. 
The Berkeley Low-cost Interplanetary Solar Sail (BLISS) project aims to utilize  small-scale technologies to create a fleet of tiny interplanetary femto-spacecraft for rapid, low-cost exploration of the inner solar system.
This paper describes the hardware required to build a $\sim$\SI{10}{\gram} spacecraft using a \SI{1}{\meter^2} solar sail steered by micro-electromechanical systems (MEMS) inchworm actuators. 
The trajectory control to a NEO, here 101955 Bennu, is detailed along with the low-level actuation control of the solar sail and the specifications of proposed onboard communication and computation.  Two other applications are also shortly considered:  sample return from dozens of Jupiter-family comets and interstellar comet rendezvous and imaging.
The paper concludes by discussing the fundamental scaling limits and future directions for steerable autonomous miniature solar sails with onboard custom computers and sensors. 


\end{abstract}






\end{frontmatter}

\section{Introduction}
\label{sec:introduction}

Recent improvements in low-cost private-sector access to space, driven in part by the CubeSat standard --- cubic modules often in \SI{10}{\centi \meter}$\times$\SI{10}{\centi \meter}$\times$\SI{10}{\centi \meter} blocks ---  have created a wide range of exciting and useful spacecraft~\cite{puig2001development,poghosyan2017cubesat}.  
Planet Labs has shown that hundreds of \SI{3}{\liter} spacecraft have  plausible commercial use-cases, and that spacecraft designed and assembled with consumer electronics standards can survive in low earth orbit (LEO).  
Manchester {\em et al.} ~\cite{manchester2013kicksat} have deployed an orbital swarm of more than one hundred ChipSats, spacecraft at the size scale of a system-on-chip (often designed in the same processes as consumer electronics) weighing only a few grams each.
The Breakthrough Starshot Initiative hopes to show that similar spacecraft can go to the nearest stars and return image data~\cite{daukantas2017breakthrough}.

The design we propose is intermediate in size to the previous successes: much smaller than a CubeSat, with greater steering capability as compared with  ChipSats and propelled by a non-consumable propulsion source by using solar sails. 
The Berkeley Low-cost Interplanetary Solar Sail (BLISS) project is intended to demonstrate that cell phone technologies and other miniaturization via technological advancements enable unprecedented capabilities in space.

The primary motivation behind small-scale solar sails is that practical, 10-100 solar sail spacecraft remain difficult to build~\cite{garwin1958solar, jack1997solar,jack2005spacefarer}.  Transit times increase linearly with sail mass loading, so any design significantly above \SI{10}{\gram \per \meter^2} leads to mission times that are on the order of dozens of years. The high cost of depositing such a sail outside the Hill Sphere of Earth is another limitation and the selection of Earth escape trajectories is also limited by sail mass loading in that most proposed trajectories require relatively fast maneuvers to maintain heading. 
At a sail mass loading of \SI{10}{\gram \per \meter \meter}, leaving Earth orbit and transiting to near-Earth objects (NEOs) both require roughly a year ~\cite{sauer1999solar,dachwald2005optimal}. Such a mass loading also allows for more agile maneuvers, as discussed in \sect{sec:control}.
Unfortunately, achieving such a low mass loading is a severe challenge for  conventional spacecraft.  
Even CubeSats, among the smallest of spacecraft, require hundreds of square meters of sail area to achieve a mass loading of \SI{10}{\gram \per \meter^2}. 
Missions like New Horizons and OSIRIS-REx had a dry mass of hundreds of kilograms and would require sail areas in the tens of thousands of square meters, the area of many football fields, to achieve the desired mass loading. 
Sails of this size require sophisticated deployment design, and the difficulty of developing low-mass sail-deployment hardware has played a role in their lack of use to date. 
The low mass  of consumer electronics enables a new class of spacecraft driven by solar radiation pressure.  These new capabilities have the potential to dramatically lower cost and decrease mission duration for inner solar system reconnaissance and sample return. Demonstration of the potential of the new small-scale spacecraft development will proceed in phases. The first science mission we propose will be to image 10s to 100s of NEOs. The second will be to collect pristine cometary materials from 1,000's of Jupiter-family comets. 
Missions to many celestial objects are within the capability of our \SI{10}{\gram} robot, but for an initial demonstration, choosing targets with orbits close to 1 astronomical unit (AU) from the Sun simplifies the design of the power and thermal systems on the spacecraft.  
There are roughly 20,000 known NEOs, roughly 1,000 of which are believed to be asteroids greater than \SI{1}{\kilo \meter} in diameter.
Only 10 of these NEOs have been visited by spacecraft.

Beyond NEO reconnaissance, there are many potential applications of BLISS in solar system exploration and planetary science. 
Swarms of approximately \SI{10}{\gram} interplanetary spacecraft would enable rapid sample return from dozens of Jupiter-family comets.  Comets contain the building blocks of the solar system, preserved in deep freeze for 4.6 billion years.  Sample return of pristine cometary material was identified as a high priority in the most recent Planetary Decadal Survey, and a comet sample return mission --- from a single comet  --- was one of two finalists in the most recent New Frontiers (\$1B-class) competition.   Even with an aggressive schedule, cometary samples would be expected no earlier than the mid-2040's with a New Frontiers approach.  However, a fleet of \SI{10}{\gram} interplanetary spacecraft has the potential to collect cometary samples from dozens of Jupiter-family comets within the next decade.   Because cometary materials are complex on the submicron scale, large samples are not required for cometary sample return:  the thousands of $\sim$10\,$\mu$m rocks and abundant organic materials would keep the cosmochemistry community busy for decades. 

The first known macroscopic interstellar object in the solar system, a ``comet'' called 1I/'Oumuamua, was discovered by PAN-STARRS in 2017. It  exhibits an extreme light curve, implying a cigar-like shape that has no parallel in any solar-system object.   'Oumuamua followed a non-Keplerian orbit, despite the lack of any detectable outgassing that could cause such a large anomalous acceleration.   These bizarre properties have even led to the suggestion that 'Oumuamua is an artificial object.   Since 'Oumuamua has not been imaged, and is on its way out of the solar system at high speed, the question is unresolved.  One additional interstellar object, 2I/Borisov, has been discovered since the discovery of 'Oumuamua, and it is thought that there on order one detectable interstellar object enters the solar system on a hyperbolic orbit per year.   Imaging of these objects is obviously of extremely high scientific interest.
A swarm of small solar sails, prepositioned in strategically-chosen heliocentric orbits, could respond rapidly to the discovery of interstellar objects, enabling high-resolution fly-by imagery.



The rest of this paper is organized as such: in section \sect{sec:spacecraft} the components needed to construct a \SI{10}{\gram} spacecraft are described.
In section \sect{sec:dynamics}, the spacecraft rotational body-centered dynamics and thrust capabilities are shown.
Section \sect{sec:mission} then presents a mission profile with the goal of acquiring $\sim$10\,cm-resolution imagery of a NEO, and section \sect{sec:control} describes the low-level control and trajectories needed for the mission.
Section \sect{sec:communications} describes communications protocols and section \sect{sec:computation} describes the associated computational requirements. Section \sect{sec:radiation} discusses the radiation implications and some workaround options for such a mission. In section \sect{sec:limits}, the ``tall poles'' that would need to be addressed for future applications, such as cometary sample return and interstellar comet rendezvous missions and the future work of creating a more compact "sporty" spacecraft. Section \sect{sec:conclusion} concludes by summarizing the overall mission and goals.

\section{The Spacecraft}
\label{sec:spacecraft}

The proposed components for the BLISS spacecraft are listed in \tab{tab:components} along with their masses and an illustration highlighting the components and their relative positions is shown in \fig{fig:sailBOM}.
None of the components listed have masses more than a few grams, and most, such as the battery and motors, are a small fraction of a gram.  

With the core of the spacecraft weighing roughly grams and costing less than \$1,000 USD, there are not many options for propulsion with the same order of magnitude in size and cost.  
Traditional chemical propulsion does not scale well to small sizes due to surface to volume issues with heat loss and low propellant burn rates \cite{krejci2018space}.  
Electric thrusters such as ion engines and electrospray engines are very promising, but are still orders of magnitude too large \cite{krejci2018space}.
Until some other technique can be reduced to a mass of grams, the best propulsion for a \SI{10}{\gram} spacecraft is a solar sail.
The fundamental limits to this scaling are discussed in \sect{sec:limits}.


\subsection{Solar Sail Sizing and Packaging}
\label{sec:sailsizing}
With this mass, a solar sail that is just over a meter in diameter is needed to accomplish a sail loading of \SI{10}{\gram \per \meter^2}.  
Aluminized mylar is available in \SI{1.5}{\micro \meter} thickness, weighs \SI{2}{\gram \per \meter^2}, and yet is strong enough to be handled, cut, folded, and thermally bonded.
A carbon fiber rod can be bent into a ring on the 
perimeter of the aluminized mylar sail and is sufficiently flexible to be bent into a multi-turn flattened disk, and sufficiently stiff to pop back to regular shape after release on orbit.
Folds of three and five would allow a square meter sail to be carried inside a 9U or perhaps 3U CubeSat for early technology demonstrations~\cite{puig2001development, toorian2008cubesat}.
In addition to the main sail, we will append a small sub-sail called the \textit{roll sail} to one edge of the main sail to achieve control authority about all body axes.
For a launch with one thousand spacecraft after the technology has been proven, the sails may be stored stacked on top of each other, with the spacecraft bodies arrayed around the perimeter.  
One thousand spacecraft of this type will weigh roughly \SI{10}{\kilo \gram}, and the stack of sails could be only a few millimeters thick.

\begin{center}
\begin{table}[t]
\centering
    \begin{tabular}{ @{} p{4cm} c @{}  }
        \toprule 
         Component & mass(g)  \\ 
         \midrule
        
         Mylar Sail & < 2 \\
         MEMS Motors + IMU & < 0.3 \\
         Optical Tx/Rx & 2 \\
         HOPG Radiator & 1 \\
         LiPo Battery & 0.33 \\
         iPhone Camera & < 1  \\
         VoCore2 PC & 2.4 \\
         Alta Solar Cells & 2 \\ 
         \midrule 
         \textbf{Total} & < 11.03 \\
         \bottomrule
     \end{tabular}
     \caption{List of Component Masses}
     \label{tab:components}
\end{table}
\end{center}

\subsection{Science Payload}
\label{sec:payload}
Imaging will be accomplished with a cell phone camera from within \SI{1}{\kilo \meter} so that the asteroid fills the field of view (from an approximately \SI{60}{\degree} field of view angle the ideal distance away to fill the aperture is $\sqrt{3}\cdot r_\text{NEO}$). This will provide  imaging with resolution on the order of \si{1}-\SI{10}{\centi \meter}, enough to differentiate regolith or other surface types on asteroids.

Communication between a spacecraft and earth will be achieved by a semiconductor laser transmitter with a diffraction-limited \SI{1}{\centi \meter} aperture, and a modestly sized single photon avalanche detector (SPAD) array with a receive aperture with a \SI{10}{\centi \meter} diameter.
This same hardware will allow peer-to-peer communication between spacecraft at more than one million kilometers, enabling a mesh network to pass data around the inner solar system.
Digital processing and storage will consist of an off-the-shelf VoCore2 embedded LINUX computer running custom software.
A micro-electromechanical system (MEMS) inertial measurement unit (IMU) will provide high-rate rotation sensing, and the CMOS camera
will provide low-rate rotation sensing, star tracking, attitude determination, and rough localization.
Propulsion will be provided by a roughly \SI{1}{\meter^2} solar sail, with MEMS motors controlling the sail and spacecraft orientation in 3 axes.  
Power will be provided by a 10$\times$\SI{10}{\centi \meter^2} primary solar panel, a small rechargeable lithium battery, and a few small solar cells distributed around the structure to maintain power during maneuvers.
Thermal control will be accomplished by control of the orientation relative to the Sun of the primary solar cells, a small solar reflector, and a highly-oriented pyrolytic graphite (HOPG) radiator fin, in addition to management of the electronics power consumption.


\begin{figure*}
    \centering
    \includegraphics[width=0.9\linewidth]{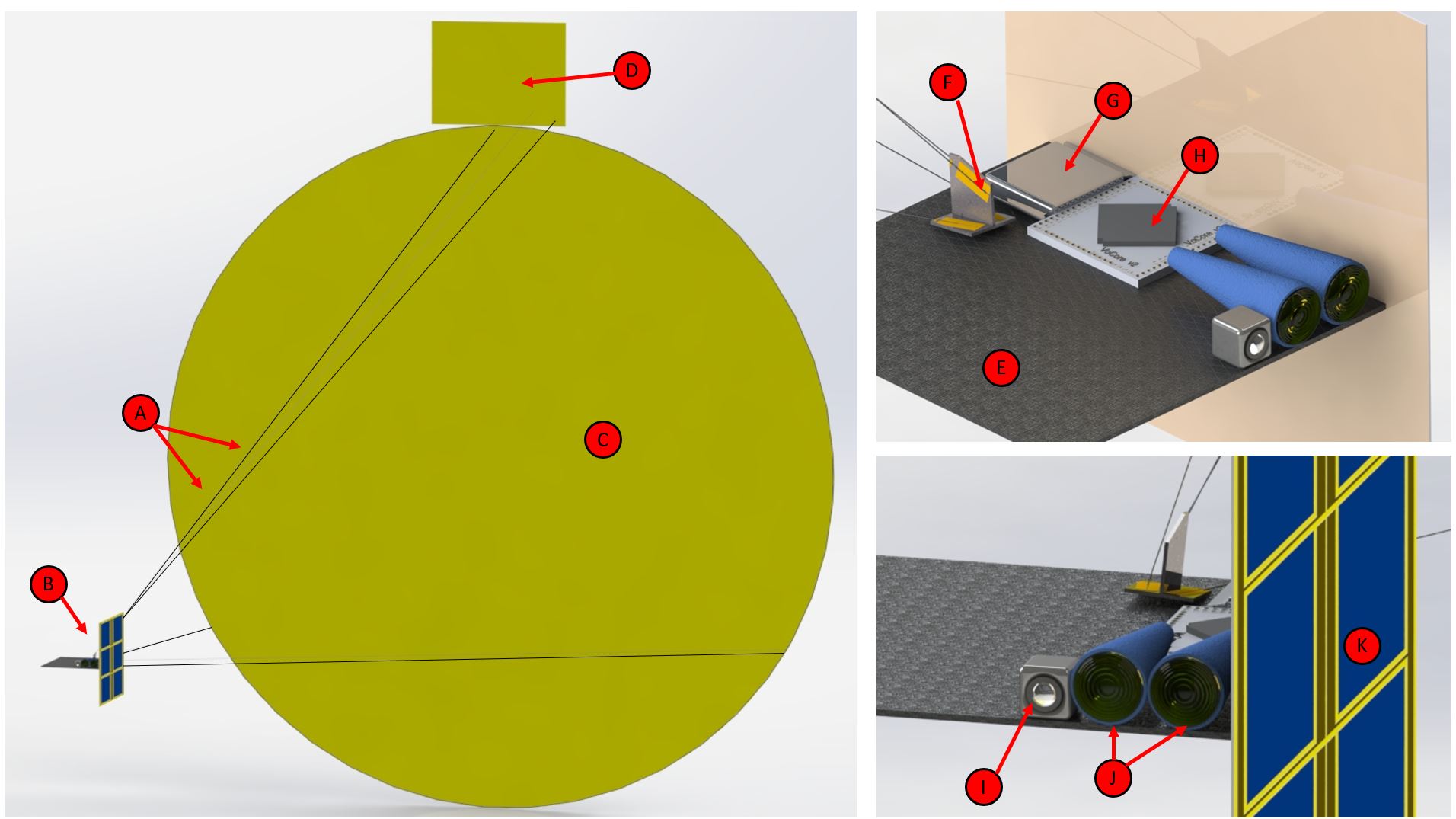}
    \caption{Rendering of the BLISS spacecraft. 
    \textit{Left}: isometric view of the solar sail craft with A) carbon fiber rods, B) spacecraft body, C) mylar main sail, and D) mylar roll sail. 
    \textit{Right-top}:spacecraft body with E) HOPG radiator fin, F) MEMS motor actuators, G) LiPo battery, and H) VoCore2 CPU. 
    \textit{Right-bottom} spacecraft body with I) IPhone camera, J) optical 853nm Tx/Rx, and K) solar panels.
    }
    \label{fig:sailBOM}
\end{figure*}
\section{Robot Dynamics}
\label{sec:dynamics}

In this section we detail the physical mechanisms that control the small solar sail through its spaceflight.

\subsection{Thrust}
\label{sec:lightpressure}
Photons have energy $E=h c/\lambda$, and momentum $p= h/\lambda=E/c$.  
A flux of photons with power $P$
being absorbed by a surface imparts a force equal to the change in
momentum, $F_\text{photon}=P/c$.  
A perfectly reflective flat sail of area $A$ at an angle $\cone$ from the solar flux of $G_\text{sc}=1.361\si{\kilo\watt\per\meter^2}$ experiences a normal force of:
\begin{equation}
    F_\textit{N} = 2 \frac{G_\textit{sc}}{c} A \cos^2(\cone) \approx (9~\mu \textit{N}/\textit{m}^2)  A \cos^2(\cone)
\end{equation}
This is the same as the gravitational attraction between the sail
and the Sun at a distance of \SI{1}{AU}
if the mass loading of the sail is roughly \SI{1.6}{\gram \per \meter^2}.
Real sails will have a reflection coefficient less than 1, some
Lambertian reflection, and pressure from black body emission from the
back side, making the peak normal force somewhat lower~\cite{mcinnes2013solar}.

The radial ($F_\textit{r}$), tangential ($F_\textit{t}$), and normal ($F_\textit{h}$), components of the force then become:
\begin{align}
    F_\textit{r} & = F_\textit{N} \cos(\cone),\\
    F_\textit{t} & = F_\textit{N} \sin(\cone)\sin(\clock),\\
    F_\textit{h} & = F_\textit{N} \sin(\cone)\cos(\clock).
\end{align}
Taking $\clock$ to be $\frac{\pi}{2}$ it can be observed that the behavior in the plane of the ecliptic gives $F_\textit{t} = F_\textit{N} \sin(\cone)$ and $F_\textit{h} = 0$.
The tangential force is maximized at an angle 
$\sin^{-1}(1/\sqrt{3}) \approx$\SI{0.6}{\radian},
giving a maximum tangential force of roughly $0.39 F_N$.
For near-circular orbits, this is the force that maximizes the rate of change of orbital energy.  
For a \SI{10}{\gram} spacecraft with a \SI{1}{\meter^2} sail at 1 AU, this maximum force corresponds to an acceleration of roughly \SI{0.3}{\milli \meter \per \second^2}.  

The sail force vector for an ideally reflective sail is always normal to the surface of a flat sail, and the radial component is always positive (away from the Sun).
Nevertheless, a solar sail can still move toward the Sun by using the tangential force to slow its orbit~\cite{garwin1958solar}.
For elliptical orbits, the desired force will vary in angle relative to the sun line.  
In general, the force in a particular direction relative to the Sun line is maximized by a sail cone angle of roughly one third of the desired angle~\cite{mcinnes2013solar}. 

\subsection{Rotation and Steering}
\label{sec:sailactuation}
\begin{figure}[h]
    \centering
    \includegraphics[width=\linewidth]{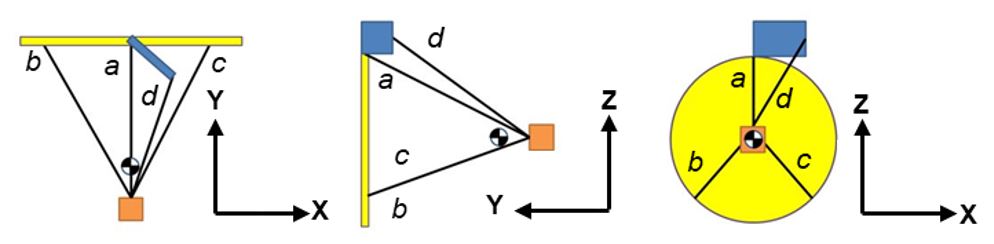}
    \caption{Illustrating the BLISS body centered reference frame and approximate center of mass. Lengths not to scale. 
    \textit{Left}: x-y plane, \textit{Middle}: y-z plane, and \textit{Right}: z-x plane.
    }
    \label{fig:body_centered_frame}
\end{figure}

Achieving the desired cone angle requires utilizing shifts in the center of mass (COM) relative to the SRP pointing vector creating a torque about the COM.
The main sail and the roll sail will be connected to the spacecraft body with four carbon fiber rods, here labeled a-d, roughly \SI{0.3}{\milli \meter} in diameter and \SI{2}{ \meter} long. 
The rods \textit{a-c} will be connected at one end to the main sail with \textit{a} designated as the stationary rod and attached at the opposing end to the main body structure. Rods \textit{b} and \textit{c} are attached at their opposing end to MEMS linear inchworm motors adjusting length to control the relative position and orientation of the main sail to the spacecraft main body. Rod \textit{d} will control the roll sail's relative motion and is affixed at one to the roll sail and the other end to MEMS motors similar to the previous two driven rods, \textit{b} and \textit{c}. The MEMS motors will be placed on the spacecraft body, as seen in \fig{fig:sailBOM} ~\cite{zoll2019mems}.  A cartoon depicting the axis and relative positions of the main sail, roll sail, carbon fiber rods, and the spacecraft body can be seen in \fig{fig:body_centered_frame}, not to scale.

Parallel and differential actuation of the first and second carbon fiber rods will move the main sail relative to the craft's payload, offsetting the COM from the SRP pointing vector providing a moment about the COM of the system, allowing control of rotation about $x$-axis (pitch) and about $z$-axis (yaw). A fourth rod, with the third motor, will control the roll sail to generate a moment about the y-axis (roll). Here, a decoupled rotation approach is chosen where the rotations about the $x$- and $z$- axes neglect the contribution of torque due to the roll flag in the calculations below. 

Coordinating movement of the inchworm motors gives an input acting on the main body of the femto-spacecraft as:
\begin{equation}
    \vec{q} = \begin{bmatrix}
    q_\text{a} & q_\text{b} & q_\text{c} & q_\roll
    \end{bmatrix}.
\end{equation}
where $q_i$ represents a linear movement, caused by the MEMS inchworm motors, of the carbon fiber rod attached to the roll sail. Whereas $q_\text{b}$ and $q_\text{c}$ represent the linear movement of rods b and c which induces a positional change on the payload relative to the center-line of the main solar sail; the axes are shown in \fig{fig:body_centered_frame}.
Such movement will create a mirrored motion of the sail relative to the origin at the COM.
Note, the sail positions ($\xsail$, $\ysail$, $\zsail$) and payload positions ($\xpayload$, $\ypayload$, $\zpayload$) are all functions of the inputs, $\vec{q}$. Still, the notation is simplified for clarity to be distances in their relative direction from the COM.

MEMS electrostatic inchworm motors have a step size that is typically designed to be on the order of \SI{1}{\micro \meter}, with a speed on the order of \SI{1}{\milli \meter \per \second} ~\cite{zoll2019mems}, and Teal {\em et al.} ~\cite{teal2021transducers} has shown a range that has been demonstrated at nearly \SI{8}{\centi \meter}. For this work, an actuation range of $\leq$$\pm$\SI{10}{\centi \meter}, a speed of \SI{1}{\milli \meter \per \second}, and a step size of \SI{2}{\micro \meter} is simulated.
The inchworm motors are powered by capacitive gap-closing actuator (GCA) arrays and push a centrally aligned spine and pawl with high frequency, electrically induced steps~\cite{rogers2006bi}.
Applied to the meter-scale shroud lines of the sail, this will generate an angular change in the sail center-line of roughly one microradian (0.2 arc seconds) per step.  
There may be significant low-frequency dynamics excited by an abrupt step, so caution in motor design is necessary.





In the body-centered frame the principal mass moments of inertia are:
\begin{align}
    \begin{split}
        \text{I}_\text{1}  & = \frac{1}{4}\sailmass \sailradius^2 + \sailmass({\ysail}^2 + {\zsail}^2) \\
        & + \text{I}_\text{mp,xx} + \payloadmass({\ypayload}^2 + {\zpayload}^2) \\
        & + \rollsailmass ({\xrollsail}^2+{\yrollsail}^2 )
    \end{split}
    \label{eq:inertialx} \\
    \begin{split}
        \text{I}_\text{2}  & = \frac{1}{2}\sailmass \sailradius^2 + \sailmass({\xsail}^2 + {\zsail}^2) \\
        & + \text{I}_\text{mp,yy} + \payloadmass(\xpayload^2 + \zpayload^2) \\
        & + \rollsailmass {\rollsailradius}^2 
    \end{split}
   \label{eq:inertialy} \\
    \begin{split}
        \text{I}_\text{3}  & =  \frac{1}{4}\sailmass \sailradius^2 + \sailmass({\xsail}^2 + {\ysail}^2) \\
        & + \text{I}_\text{mp,zz} + \payloadmass(\xpayload^2 + \ypayload^2) \\
        & + \rollsailmass ({\xrollsail}^2+{\yrollsail}^2) \\
        & + \text{I}_\text{zz,$\phi$}
   \end{split}
   \label{eq:inertialz}
\end{align}
 where $I_\text{mp,xx}$, $I_\text{mp,yy}$, and $I_\text{mp,zz}$ are the inertial moment of mass for the spacecraft body COM used to indicated positions of components within the spacecraft as related to the parallel axis theorem in the $\hat{x}$, $\hat{y}$, and $\hat{z}$ body-centered coordinate system. 
 Other cross terms exist and are negligible without loss of generality, but for early experiments these represent the principal moments of inertia used in \sect{sec:lowlevel}.


If the sail normal is moved about \SI{1}{\milli\meter} from the COM in the $\hat{x}$-direction, the resulting pitch or yaw moment will be a torque of around \SI{60}{\nano N \meter}, and an angular acceleration of:
\begin{equation}
  \ddot{\cone} = \frac{\tau_\text{z}}{I_\text{zz}},
\end{equation}
 roughly $1 \times 10^{-6}$\SI{}{\radian \per \second^2} if we take into account only the main sail giving a rough estimate for $I_\text{zz} \approx I_\text{ss,zz}$. 
 Note, due to the symmetry shown in \eq{eq:inertialx} and \eq{eq:inertialz}, the rotation about the $\hat{z}$ axis behaves similarly to the rotation about the $\hat{x}$ axis.
 Rotations about the $\hat{y}$ axis are substantially slower due to lower peak torque from the smaller roll sail, but motions in this axis are primarily for thermal regulation and data collection. These present a much smaller set of problems with much longer timescales of motion than when compared to maneuvers needed to escape geosynchronous orbit (GEO).
 


\section{Mission profile}
\label{sec:mission}
The mission profile entails orientation following an en-mass launch, spiraling out from GEO, interplanetary trajectory optimization, NEO approach, imaging, and return to Earth for data transmission. Assuming that the spacecraft will be released near GEO, atmospheric drag will be negligible, as any orbit with a perigee above roughly \SI{500}{\kilo \meter} is high enough that the energy gained per orbit from the sail is greater than the energy lost due to atmospheric drag. The months spent in transit through the Van Allen belts present a serious hazard to the off-the-shelf electronics, for more on this, see \sect{sec:radiation}. 
The return to Earth mirrors the procedure to the interstellar object.

\subsection{De-tumble and Orient} 
After release from the launch vehicle, the spacecraft will need to kill any residual rotation imparted by release.
Orientation control of the spacecraft is provided by MEMS inchworm motors pulling on shroud lines on the sail. 
For rotation rates above \SI{1}{\degree \per \second}, the on-board IMU  will inform the control system's decisions.  
Low-cost MEMS gyros have thermal-noise-limited resolution and bias drift of a small fraction of a degree per second.
Below that rate, the camera will be used in a time lapse exposure to find the axis of rotation and rotation rate.
Once the rotation rate is sufficiently low, a standard "lost in space" algorithm will be run on the star images to accurately determine orientation~\cite{liebe1993pattern}.
If there is sufficiently low magnitude and duration of tumbling on release, the IMU itself may be able to provide fairly accurate initial orientation estimates.
During this initial phase, communication between spacecraft and the launch vehicle will be possible to a distance of a few hundred meters using the integrated WiFi radio available in almost all embedded LINUX platforms.

\subsection{Spiral out}
Once oriented, the spacecraft will begin the process of increasing its orbital altitude until it crosses out of the Earth's sphere of influence, utilizing a trajectory in the Sun-Earth orbital plane.  
A conservative estimate for escape time is given by~\cite{mcinnes2013solar}:
\begin{equation}
    t_\text{escape} = \frac{2800 ~\text{days}}{\beta \sqrt{6371 + h}},
\end{equation}
where $h$ is the initial orbital altitude in km, and $\beta$ is the sail lightness number, the ratio of the maximum light pressure acceleration to the solar gravitation, which to first order is independent of distance from the Sun. 
For a sail mass loading of \SI{10}{\gram \per \meter^2}, $\beta\approx 0.16$.
Starting from GEO this process takes about three months.

Imaging the Earth, moons, and stars provides the necessary position input for the attitude control system to calculate the necessary sail angle during this process.
The chosen in-plane spiral-out trajectory is one adapted from McInnes \cite{mcinnes2013solar}, where $\alpha$ traverses from \SI{0}{\degree} to \SI{90}{\degree} monotonically with the mean anomaly ($\theta$), switches relatively quickly from \SI{+90}{\degree} to \SI{-90}{\degree} at the 'top' of the orbit, and then proceeds monotonically sweeping from \SI{-90}{\degree} to \SI{0}{\degree} throughout the course of each orbit  \sect{sec:control}.
During this phase of the mission, the spacecraft can be in periodic communication with ground stations on Earth using its primary optical communication system.
In addition, the BLISS system can establish communication with its neighboring BLISS partner spacecraft (BliPS).

\subsection{Localization and Matching Orbits}
There is a wide variety of NEO orbits, the majority of which will take between a few months and two years to match given our sail mass loading~\cite{chodas2015overview}  
During this time, the spacecraft may be in communication with many of its neighbors, and able to use a combination of triangulation and optical time-of-flight (TOF) to determine its location in the solar system.
Round trip TOF with a 1 Mbps optical communication system can easily achieve range errors of less than a kilometer at distances 
of at least one million kilometers, and small fractions of that error are possible with specialized hardware development.
Pointing accuracy in the communication system should provide angular position measurements of roughly 0.1 mrad, or \SI{100}{\kilo \meter} lateral accuracy at \SI{1E6}{\kilo \meter} range, with errors accumulating the more hops the spacecraft is from a known reference point.
If a fleet of spacecraft leaves the range in which communication with Earth is possible, it may have a reasonably accurate picture of the relative position of each member of the fleet (kilometers of error or less), but a very poor estimate of where the fleet is in the solar system (one hundred thousand kilometers).

For those spacecraft without communication to their neighbors, they will rely on triangulation using the location of the planets against the stellar background.
Unmodified cell phone cameras are able to take pictures of stars down to visual magnitude 5 through Earth's atmosphere.  With minimal effort the inner six planets can be located in these images with an accuracy of roughly one milliradian.  
This will allow a spacecraft to localize itself in interplanetary space to within a few hundred thousand kilometers at a single point in time using several photographs.  With better image processing algorithms and fusion of sensor data taken over many days, it is likely that this accuracy will improve by at least one order of magnitude.  
In either case, once it is close to the target NEO the spacecraft will use its camera to do terminal guidance.

Asteroids of 1-2 km diameter have absolute magnitudes, $H$, in the range of 15 to 17.  
The apparent magnitude, $m$ is given by, 

\begin{equation}
m =  H + 5 \text{log}\frac{d_\text{BS}d_\text{BO}}{(1 \text{AU})^2}-2.5\text{log}(\text{$\Phi$($\gamma$)}).
\end{equation}

Where $d_\text{BO}$ is the distance between the body and observer, $d_\text{BS}$ is the distance between the body and Sun, and $d_\text{OS}$ is the distance between the observer and Sun. Here $\gamma$ is the angle between the body-Sun and body-observer lines, expressed as 
\begin{equation}
    cos(\gamma)=\frac{d_\text{BO}^2+d_\text{BS}^2-d_\text{OS}^2}{2d_\text{BO}d_\text{BS}}
\end{equation}
and $\Phi(\gamma)$ is the phase integral which accounts for the reflected light and is a number between 0 and 1 given by the approximation in Whitmell \cite{whitmell1907brightness}.
At 0.01 AU distance from the spacecraft, a 1 to 2 km diameter asteroid 1 AU from the sun will have an apparent magnitude
of 5 to 7.  So the spacecraft should be able to get close enough to a large asteroid
by navigating based on sightings of the planets to be able to see the asteroid with the camera.  


\subsection{NEO Approach, Image, and Return}
 The mission objective of asteroid imaging requires a down-selection to decide on an asteroid for case study that can act as a representative example for the benefits of the BLISS system in operation. As such, selection will be narrowed to include only near-Earth asteroids with diameters of 1 to 2 km. There are roughly 400 asteroids of 1 to 2 km diameter in near-Earth orbits ~\cite{chodas2015overview, chamberlin2010jpl}. Of these, the the 101955 Bennu asteroid is taken as a prime example of state-of-the-art asteroid rendezvous recently accomplished. Images will be taken of Bennu from a parking orbit of \si{1}-\SI{3}{\kilo \meter} from the surface. After the spacecraft has determined that it has sufficient images of the NEO, it will begin the return trip to Earth, following inverse order outlined for the trajectories specified in this paper. After the return trip, BLISS will transmit its data down to receivers on Earth from GEO.

\section{Control and Planning}
\label{sec:control}
Control of a solar sail spacecraft is a well studied problem in the context of orbit~\cite{morrow2001solar} and optimal trajectory control~\cite{mengali2005optimal}.
In this section we detail the control laws deployed to utilize MEMS inchworm motors for low-level control planning towards NEO rendezvous. 

\subsection{Low-level Control}
\label{sec:lowlevel}
The discrete steps of the MEMS inchworm motors provide precise, reliable actuation of the control rods which determine the pointing vector of the BLISS spacecraft.
There will be two separate low-level control systems to maintain the robot's orientation in space: a pitch and yaw controller actuating the main sail and a roll controller actuating the auxiliary sail.
Differential application of the two main control rods induces a desired linear shift of the force vector relative to the COM causing a pitch, roll, and yaw rotation following proportional-derivative (PD) control. 
Defining a local state of the system, $\vec{s}$ as:
\begin{equation}
    \vec{s} = \begin{bmatrix} 
    $x$ & $y$ & $z$ & \roll
    \end{bmatrix},
    \label{eq:lowlevelaction}
\end{equation}
the low-level controller will follow the desired state command from the trajectory planner outlined in \sect{sec:interplanetarytrajectory}.
The four-dimensional control input, defined in \sect{sec:sailactuation}, controls the attitude of the spacecraft. Given error from a desired orientation $e_i$ (e.g. $e_\cone = \cone_d - \cone$), PD control is defined by
\begin{equation}
    u_d = K_P\cdot e_i + K_D\cdot \dot{e}_i.
    \label{eq:pid}
\end{equation}

\begin{equation}
    \epsilon_\text{attitude} \geq \|e_{\cone} \|.
\end{equation}

Initial selection of the PD control parameters is accomplished using standard linear control theory methods to get $\text{K}_\text{P}$ and $\text{K}_\text{D}$ initial values for simulation for the trajectories discussed in \sect{sec:interplanetarytrajectory} and then hand tuning is performed to get higher order accuracy.  

\subsection{Earth Orbit Maneuvers}
\label{sec:geo}
\begin{figure}
    \includegraphics[width=0.4\textwidth]{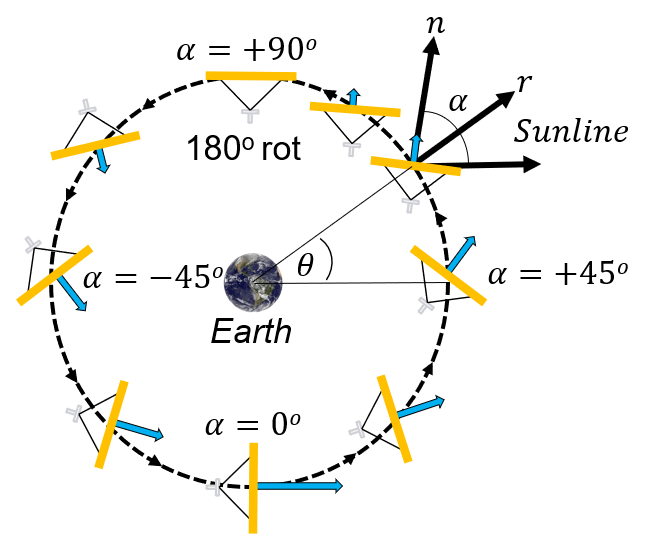}
    \caption{McInnes orbit rate steering law showing direction of sail normal throughout a full circular orbit around Earth. Adapted from \cite{mcinnes2013solar}. Not to scale.}
    \label{fig:McInnes_diagram}
\end{figure}
The modified orbit rate steering law (MORSL) will be used to accomplish Earth escape, which is adapted from McInnes \cite{mcinnes2013solar}. In the original McInnes orbit rate steering law (OgMORSL), described as an orbit in the plane of the ecliptic, the spacecraft cone angle follows monotonically the mean anomaly in a counterclockwise rotation about Earth starting at the 3 o'clock position in a $\SI{45}{\degree}$ cone angle, where 12 o'clock is the direction of Earth orbit in the projected heliocentric coordinate system. At the 12 o'clock position, the spacecraft flips orientation, presumably negligibly, swiftly such that it is relatively instantaneous and thereafter continues its cone angle rotation again matching monotonically to mean anomaly. The crucial low-level control maneuver for a solar sail orbiting around Earth in the plane of the ecliptic is the rapid switch of orientation to maintain the required orientation of the reflective side of the sail facing the Sun, assuming the 3 o'clock position is directly in the Earth's shadow as the Sun will be at the 9 o'clock position.
Ideally, upon reaching the apex of an orbit, an ideal solar sail must switch from a cone angle of $+\SI{90}{\degree}$ to \SI{-90}{\degree}\cite{mcinnes2013solar} as indicated in \fig{fig:McInnes_diagram}. 

\begin{table}[t]
\centering
\label{table:earthIC}
\begin{tabular}{@{}llll@{}} \toprule
ICs & Value & Units \\ \midrule
\ \ \ $\cone$            & 0.8613 & rad       \\
\ \ \ $\dot{\cone}$      & 3.6459 $\times 10^-5 $ &  rad \SI{}{\per \second} \\
\ \ \ $r$                & 42164 & \SI{}{\kilo \meter}      \\
\ \ \ $\dot{r}$          & 0.0000 & \SI{}{\kilo \meter \per \second}      \\
\ \ \ $\theta$           & 0.1519 & rad      \\
\ \ \ $\dot{\theta}$     & 7.2919 $\times 10^-5 $ & rad \SI{}{\per \second}  \\   \bottomrule 
\end{tabular}
\caption{Initial conditions used for the Earth escape trajectory.}
\end{table}

\begin{figure}[ht]
\centering
    \begin{subfigure}{0.4\textwidth}  
        \centering
        \includegraphics[width=1.0\textwidth]{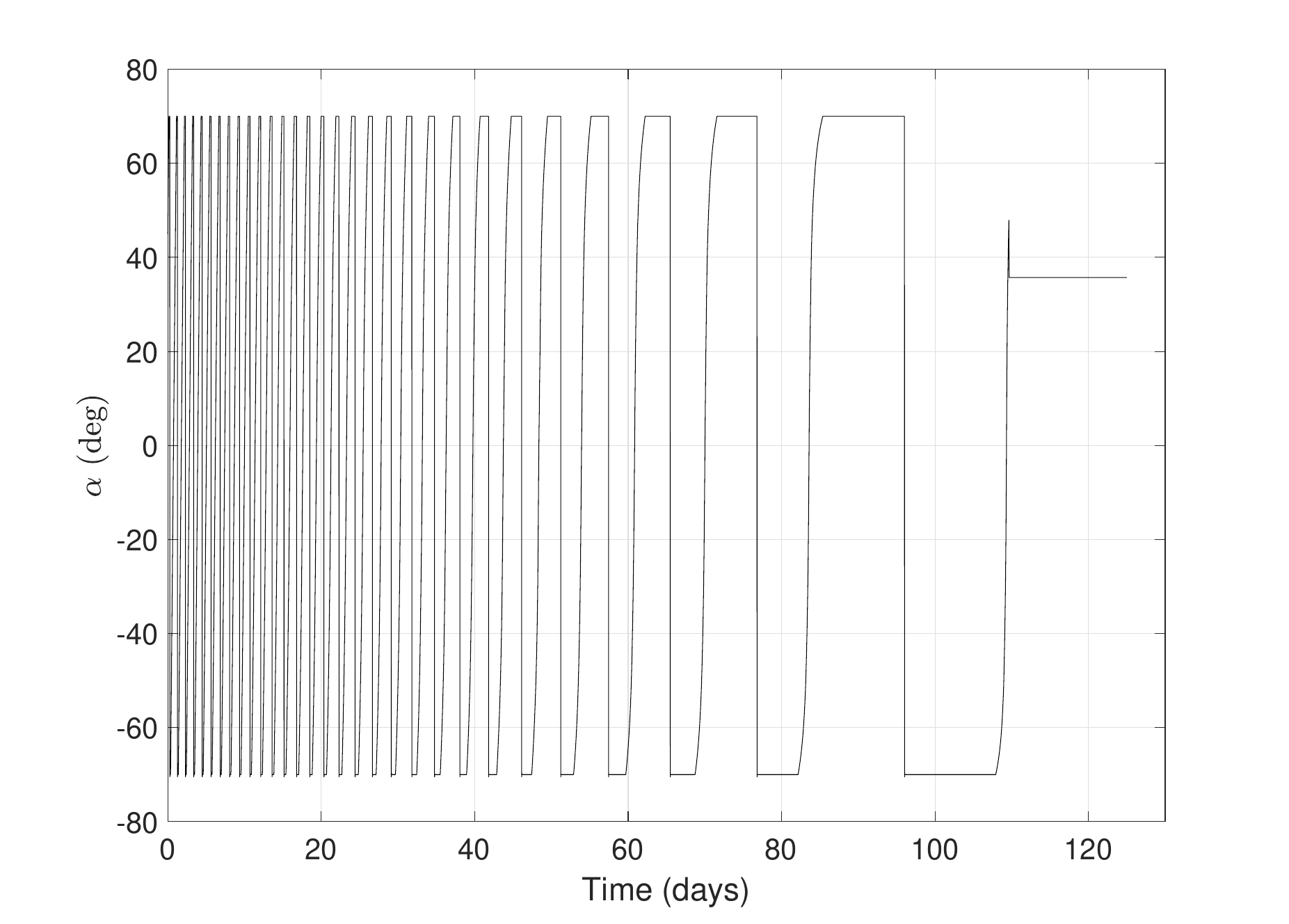}
        \caption{Spiral out cone angle.}
        \label{fig:cone_angle_full}
    \end{subfigure} 
    ~
    \begin{subfigure}{0.4\textwidth}  
        \centering
        \includegraphics[width=1.0\textwidth]{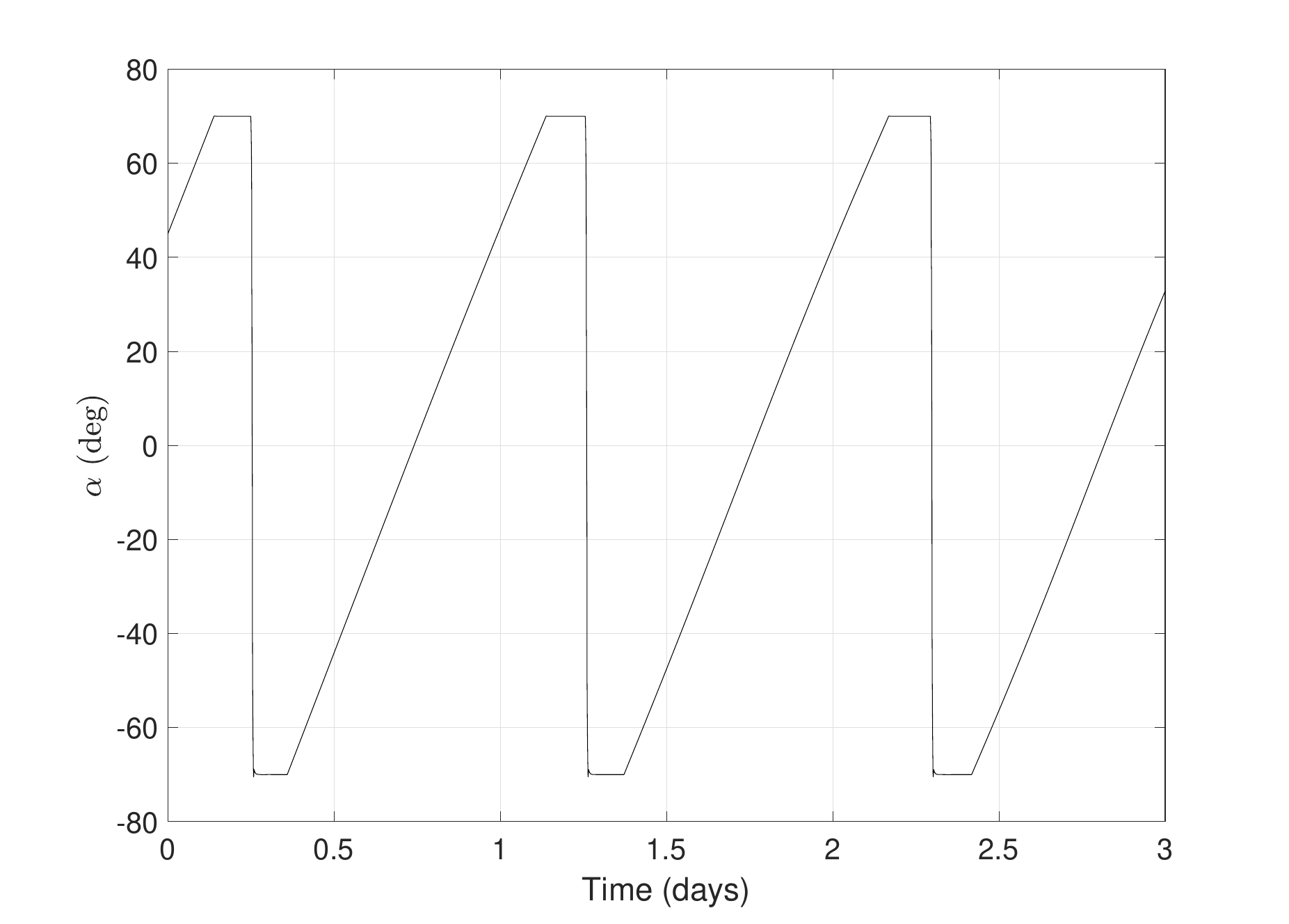}
        \caption{Spiral out cone angle (3days).}
        \label{fig:cone_angle_3day}
    \end{subfigure}
    \centering

    \caption{Simulated real cone angle for a) 125 days and b) 3 days.}         
    \centering

    \label{fig:cone_angle}
\end{figure}

Given constraints on the angular acceleration of the spacecraft body outlined in \sect{sec:sailactuation} and the drop-off in force on a solar sail with the sine of the cone angle, as it reaches $\pm\SI{90}{\degree}$ will approach zero. As such, the cone angle is limited so as to maintain fine control of the trajectory. As the cone angle reaches $\pm\SI{70}{\degree}$ the SRP force nears $\SI{10}{\percent}$ of its nominal value and represents a chosen saturation limit. Additionally, the swift switch from $+\SI{70}{\degree}$ to \SI{-70}{\degree} at the 'top' of the Earth orbit represents a crucial maneuver to managing orbital energy losses over the escape trajectory. These losses, if unchecked result in an overall slower orbit-raising trajectory both increasing time to escape and also the likelihood of an undesirable early touchdown on the surface of the Earth. 

\begin{figure*}[t] 
\centering
        \includegraphics[width=0.55\linewidth]{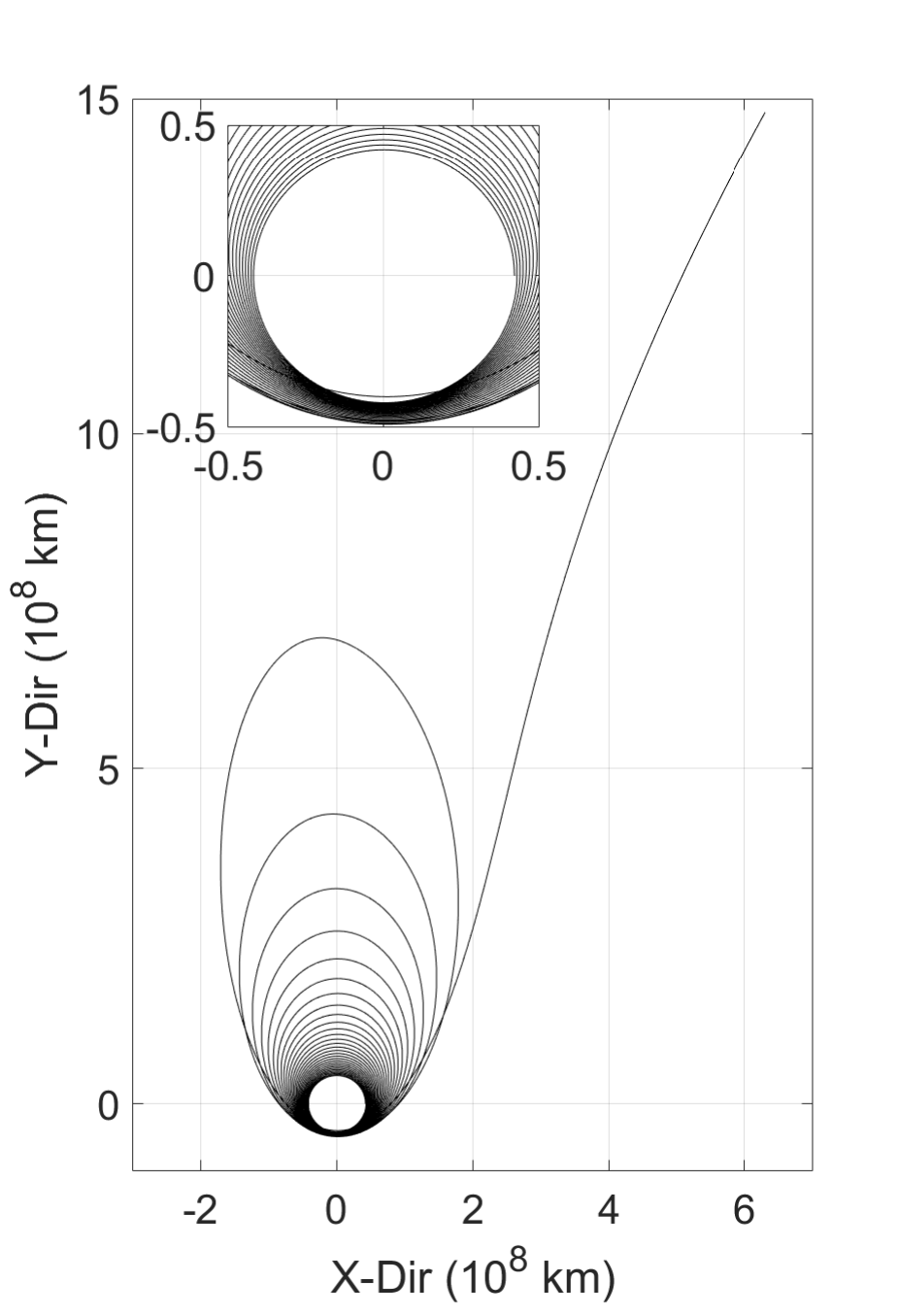}
        \caption{Spiral out XY-trajectory.}
        \label{fig:earth_orbit_full}
\centering
\end{figure*}

The downstream effect of inserting these practical non-idealities is an initially less rapid spiraling out of Earth orbit for the craft and a more eccentric orbit than the predicted OgMORSL which gains angular speed as the spacecraft spirals in towards the Earth on each successive orbit. Overall, this maneuver allows the spacecraft using the MORSL to trail the OgMORSL orbit by nearly 47 days, a relatively minor loss in performance that is overcome once leaving the Earth's gravitational pull \fig{fig:earth_orbit_full}.

\begin{figure*}
\centering
    \begin{subfigure}{0.55\linewidth} 
        \includegraphics[width=1\linewidth]{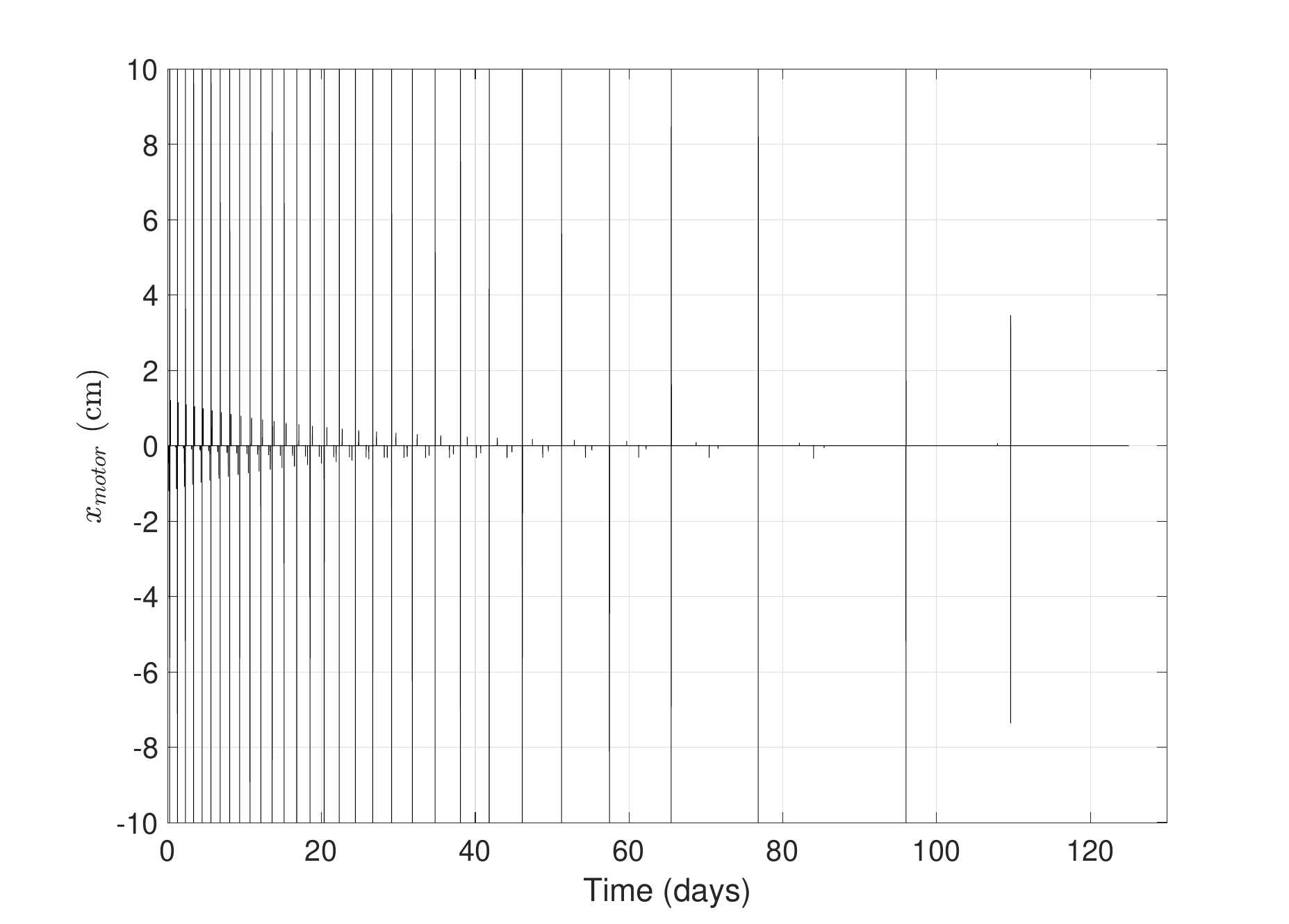}
        \caption{Motor steps along 125 day trajectory.}
        \label{fig:x_motor_125day}
    \end{subfigure}
    ~
    \begin{subfigure}{0.43\linewidth} 
        \includegraphics[width=1\linewidth]{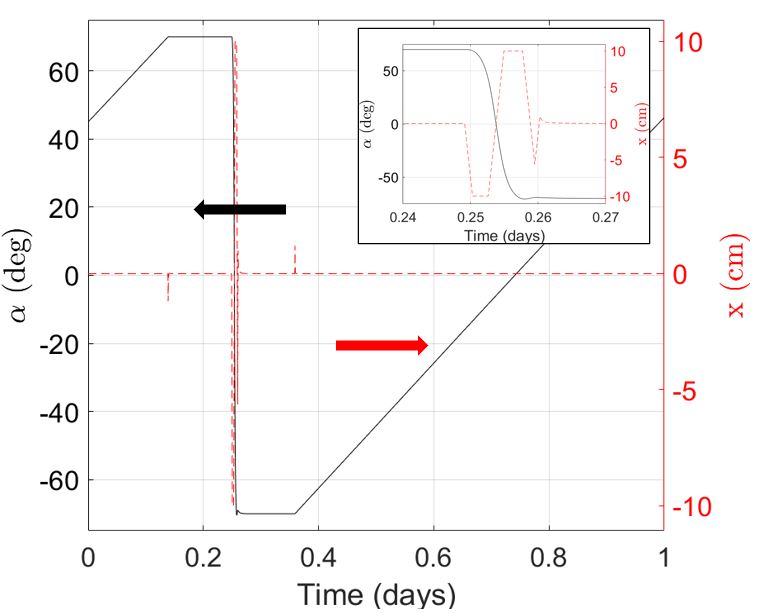}
        \caption{Motor steps and cone angle along 1 day trajectory.}
        \label{fig:cone_angle_and_x_motor}
    \end{subfigure}
    \caption{a) Cone angle for the full trajectory and b) 1 day trajectory of cone angle (left y-axis) and motor steps of MEMS inchworm (right y-axis).}
    
\centering
\end{figure*}

\begin{figure*}[h]
    \includegraphics[width=\linewidth]{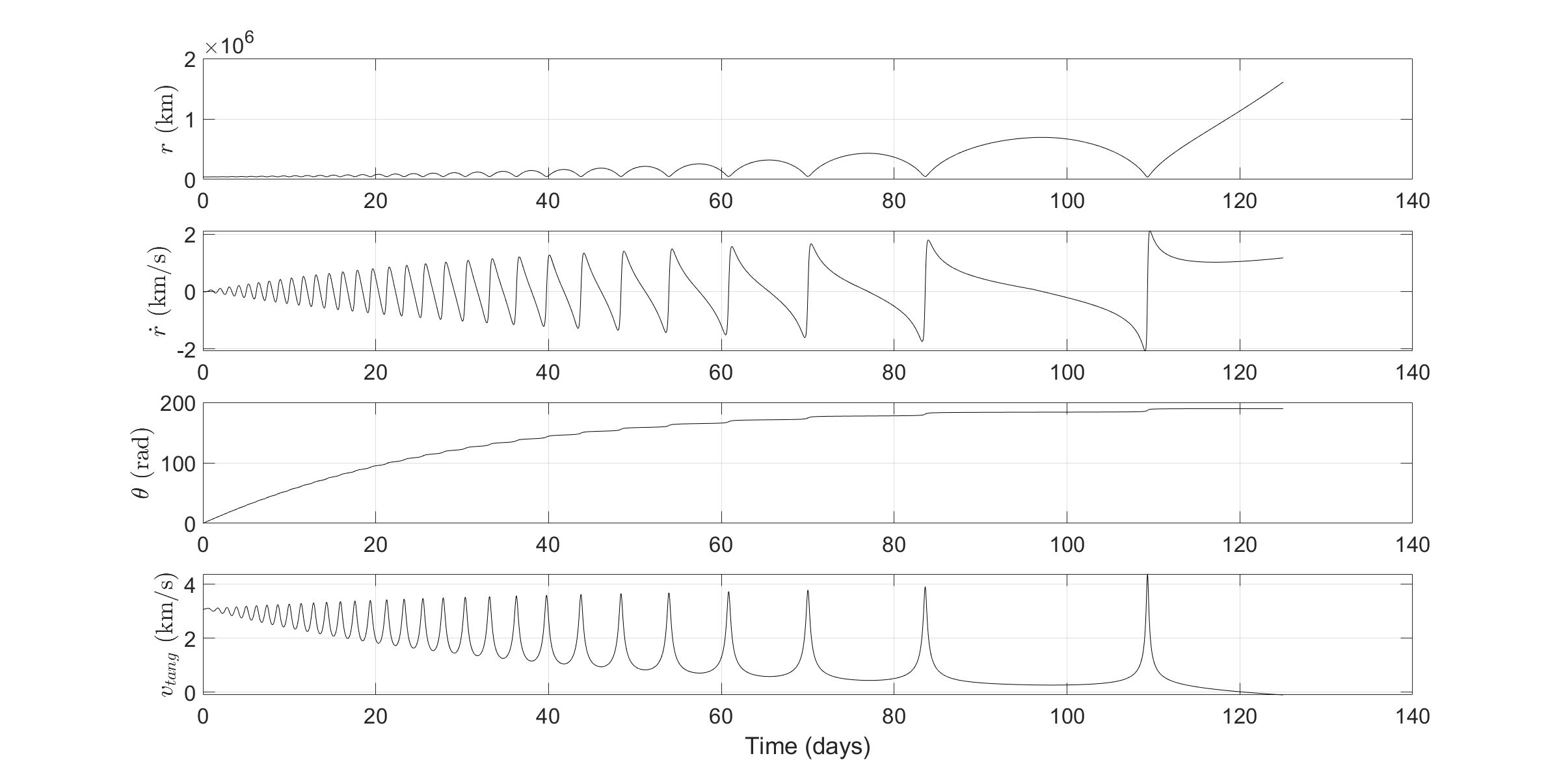}
    \caption{Orbital parameters for Earth escape trajectory (125 days). Here r is radial distance from Earth, $\dot{r}$ is radial speed, $\theta$ is mean anomaly about Earth, and $\dot{\theta}$ is angular speed.}
    \label{fig:parameters_125day}
\end{figure*}

Future work will document the energy losses and other considerations when operating at different levels of Earth's orbit, including drag induced by diffuse particles of air in the upper exosphere, utilizing the tidal perturbations of the moon for a slingshot maneuver, singularities in the 3D case, trajectories utilizing Lissajous curves, and the effect of the Van Allen radiation belt on electrical and navigation performance including error correction operations.


\subsection{Intercept Trajectories}
\label{sec:interplanetarytrajectory}
Here, treatment of the two-stage approach to the three-dimensional, near-Earth object (NEO) rendezvous problem is considered. 
In the first stage, the spacecraft matches the out-of-plane motion of the spacecraft, namely the inclination and longitude of the ascending node, to that of the NEO. 
A maximum inclination (and ascending node) rate-of-change steering law, as described in \cite{mcinnes2013solar}, may be used in this initial stage. 
Once the orbital plane change maneuver is complete, synchronization is accomplished to the in-plane motion in the second stage. 
The following trajectory optimization problem is formulated to achieve in-plane rendezvous with the NEO. 
Since solar sails do not require fuel or energy to perform orbital maneuvers, a time-optimal objective \eq{eqn:objective} is used, where the time of rendezvous, $t_f$, is a decision variable. 
Furthermore, due to the decoupling of out-of-plane and in-plane motion, one may represent the spacecraft in this second-stage with a simpler four-state system expressed in a polar coordinate system \eq{eqn:dynamics}. 
The radial position $r$ from the Sun's center, radial velocity $u$, and tangential velocity $v$ may be normalized with respect to initial conditions \eq{eqn:initialconditions}, and the angular position $\theta$ is with respect to a reference horizontal. Assuming a constant normalized peak solar sail specific force $F$ for the spacecraft in the problem formulation. The rendezvous condition of matching positions is captured by \eq{eqn:rendezvousconstraint} where there is enforcement of the terminal Eucldiean distance, expressed in Cartesian coordinates, between spacecraft and NEO to be within a tolerance value $\epsilon_{xy} = 1e\text{-}5$. 
The box constraints \eq{eqn:terminalconstraints} ensure that the rendezvous condition of matching velocities is satisfied with given tolerances, allowing the spacecraft to maintain the same orbital trajectory as that of the NEO. 
Note the assumed known state trajectory for the NEO in both polar and Cartesian coordinates, as represented by the overbar notation. 
Finally, control authority is derived from the cone angle $\cone$, the angle between the sail normal and the sun-sail line. The cone angle is bounded to be within $\pm 70^\circ$ to avoid the loss of control authority at $\pm 90^\circ$.

\begin{align}
&\minimize_{ \cone, t_{f} } \hspace{1.0cm} \int_{t_0}^{t_{f}} 1 \ dt  \label{eqn:objective} \\
  &\ \text{subject to}  \nonumber \\
  &\begin{bmatrix} \dot{r}(t) \\ \dot{u}(t) \\ \dot{v}(t) \\ \dot{\theta}(t) \end{bmatrix} = \begin{bmatrix} u(t) \\ \frac{v^2(t)}{r(t)} - \frac{1}{r^2(t)} + \frac{\hat{F}}{r^2(t)} \cos^3(\cone(t)) \\ -\frac{u(t) v(t)}{r(t)} + \frac{\hat{F}}{r^2(t)} \cos^2(\cone(t)) \sin(\cone(t))  \\ \frac{v(t)}{r(t)} \end{bmatrix} \nonumber \\ 
  &\hspace{4.5cm} \forall \ t \in [t_0, \ t_f] \label{eqn:dynamics} \\
  & [r(t_0) \ u(t_0) \ v(t_0) \ \theta(t_0)]^\top = [r_0 \ \ u_0 \ \ v_0 \ \ \theta_0]^\top \label{eqn:initialconditions} 
\end{align} 
\begin{align}
  \left\lVert 
  \begin{bmatrix} r(t_f) \cos\big(\theta(t_f)\big) - \bar{x}_{\scriptscriptstyle NEO}(t_f) \\ 
  r(t_f) \sin\big(\theta(t_f)\big) - \bar{y}_{\scriptscriptstyle NEO}(t_f)\end{bmatrix} 
  \right\rVert_2 & \leq \epsilon_{xy} \label{eqn:rendezvousconstraint} \\
  \left\lvert 
  \begin{bmatrix} r(t_f) \\ u(t_f) \\ v(t_f) \end{bmatrix} - \begin{bmatrix} \bar{r}_{NEO}(t_f) \\ \bar{u}_{NEO}(t_f) \\ \bar{v}_{NEO}(t_f) \end{bmatrix} 
  \right\rvert & \leq \begin{bmatrix} \epsilon_{r} \\ 
  \epsilon_{u} \\ \epsilon_{v} \end{bmatrix}  \label{eqn:terminalconstraints} \\
  \lvert \cone(t) \rvert \leq \alpha_{max} \hspace{1.0cm} & \forall \ t \in [t_0, \ t_f] \label{eqn:controlbound}
\end{align}
This problem formulation with nonlinear dynamics and non-convex constraints is solved using recent advances in sequential convex programming \cite{Szmuk}. 
Although a single approach that simultaneously matches out-of-plane and in-plane motion may result in more time-optimal solutions, here it is illustrated that the problem may be decoupled. 
Future work will focus on a unified, single-stage approach to the rendezvous problem. 

This numerical example uses the problem parameters listed in \tab{table:params}, where $\mu$ is the standard gravitational parameter of the Sun and $r_0$ is a reference distance. 
At this reference distance a specific value is assumed for the maximum solar radiation pressure, $P$. 
Assuming a satellite mass, $m$, and solar sail reference area, $\sailarea$, the computed normalized peak solar sail specific force is $\hat{F} = \frac{P\sailarea}{m (\mu/r_0)^2}$.

\begin{table}[t]
\centering
\begin{tabular}{@{}llll@{}} \toprule
Parameter & Value & Units \\ \midrule
\ \ \ $\mu$            & 1.327$\times$10$^{20}$ & \SI{}{\meter^3 \per \second^2}       \\
\ \ \ $r_0$             & 1.495$\times$10$^{11}$ & \SI{}{\meter}      \\
\ \ \ $P$               & 9.08$\times$10$^{-6}$ & \SI{}{\newton \per \meter^2}      \\
\ \ \ $m$               & 0.01 & \SI{}{\kilo \gram}      \\
\ \ \ $\sailarea$               & 1.0 & \SI{}{\meter}  \\  
\ \ \ $\hat{F}$               & 0.1531 &   \\ \bottomrule 
\end{tabular}
\caption{Parameters used for the trajectory optimization to 101955 Bennnu.}
\label{table:params}
\vspace{15pt}
\centering
\begin{tabular}{@{}lllll@{}} \toprule
\ \ Body & $\hat{e}_1$ (AU) & $\hat{e}_2$ (AU) & $v_1$ (\SI{}{\kilo \meter / \second}) & $v_2$ (\SI{}{\kilo \meter / \second}) \\ \midrule
\ \ \ NEO           & 0.0018 & \phantom{-}0.9027 & -34.0904 & \phantom{2}2.1930 \\
\ \ \ Sail             & 0.9744 & -0.2543 & \phantom{-3}7.0391 & 28.7254  \\ \bottomrule 
\end{tabular}
\caption{Initial conditions for trajectory to 101955 Bennnu.}
\label{table:ics}
\end{table}

The initial conditions of the solar sail and NEO are listed in Table \ref{table:ics}, where the orbital positions and velocities of Earth and asteroid 101955 Bennnu are assumed (projected on the Earth's ecliptic plane) at the epoch of September 8, 2016. 
In this preliminary study, it is assumed that the solar sail starts in the Earth's nearly circular orbit about the Sun.

\begin{figure*}[t]
\centering
\begin{center} 
    \small{\cblock{0}{0}{0} Solar Sail (\textcolor[rgb]{.0,.0,.0}{---})\quad
    \cblock{200}{20}{20} NEO (\textcolor[rgb]{.8,.1,1}{- -}) \vspace{-10pt} 
    }\end{center} 
     \begin{subfigure}[b]{0.26\textwidth}  
       \centering
        \includegraphics[width=1.0\textwidth]{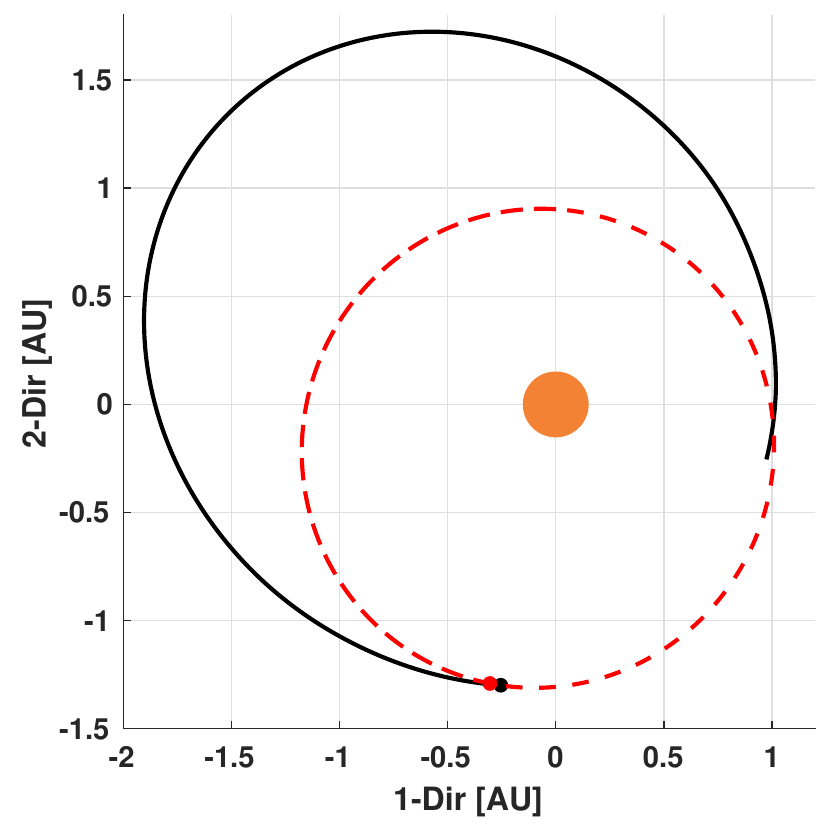}
        \caption{Orbital trajectories of solar-sail and NEO about the Sun.}
        \label{fig:inplane}
    \end{subfigure} 
    ~
    \begin{subfigure}[b]{0.34\textwidth}
        \centering
        \includegraphics[width=1.0\textwidth]{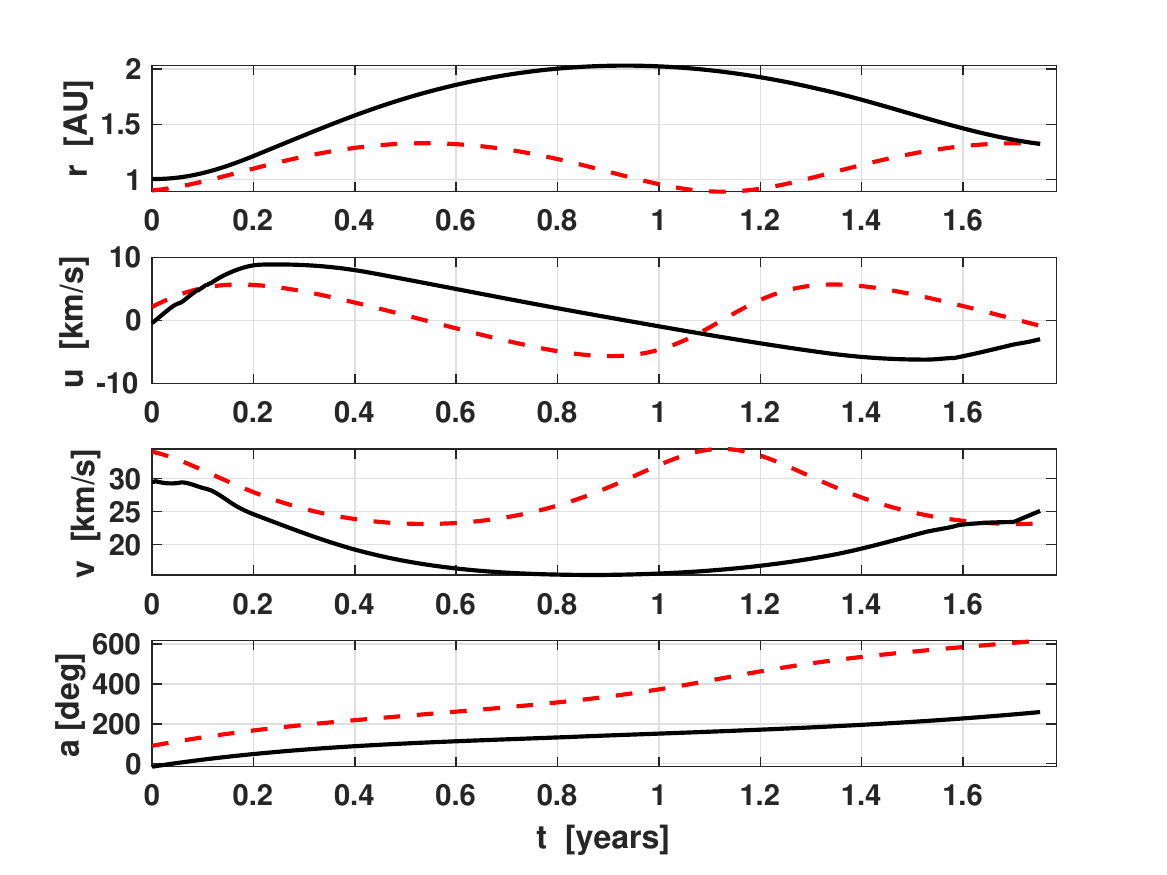}
        \caption{Optimal solar-sail state trajectory with respect to NEO state trajectory.}
        \label{fig:state}
    \end{subfigure} 
    ~
    \begin{subfigure}[b]{0.34\textwidth}  
        \centering
        \includegraphics[width=1.0\textwidth]{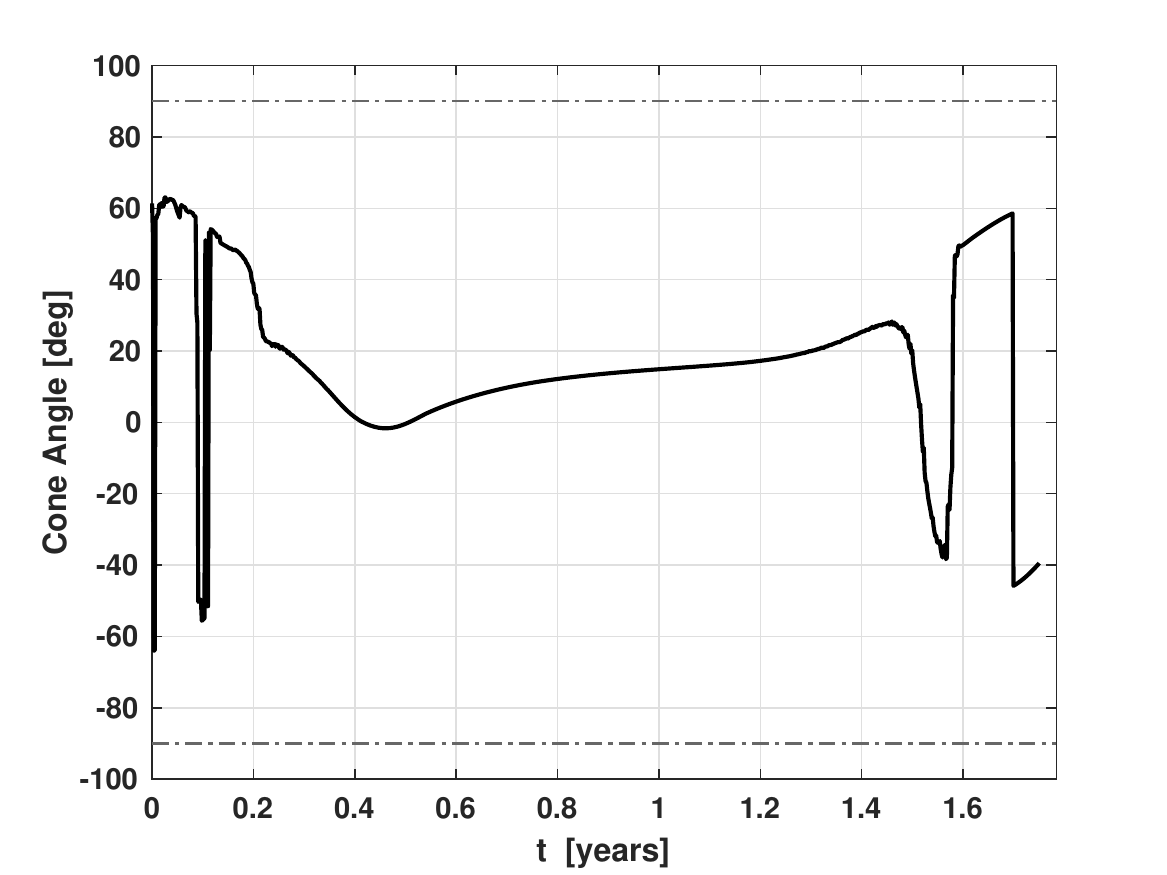}
        \caption{Optimal input trajectory and $\pm$90 degree input bounds. }
        \label{fig:input}
    \end{subfigure} 
    \label{fig:traj}
  \caption{Simulated, time-optimal trajectory of solar-sail to rendezvous with NEO is shown in shown in \fig{fig:inplane}. The initial conditions of the planar problem are listed in Table \ref{table:ics}. The states for the solar-sail and the NEO are shown in \fig{fig:state} and the optimal cone angle to achieve it is shown in\fig{fig:input}.}
\end{figure*}

As shown in \fig{fig:inplane}, our trajectory optimization problem finds a solution where the solar sail (solid, black trajectory) successfully completes a rendezvous with the NEO (dashed, red trajectory). 
Note that the solar sail's minimum-time strategy is to push radially outward, slowing down in speed, before returning towards the periapsis of its now highly elliptical orbit. 
Rather than attempting to ``catch up'' and close the angular displacement between the solar sail and the NEO, the solar sail effectively waits for the much faster NEO to complete a full orbit before rendezvous.

In \fig{fig:state}, it is observed that the optimal radial position and velocity of the solar sail match closely to those of the NEO at the end of the trajectory, signifying rendezvous. 
The duration of the maneuver is approximately 640 days (1.75 years), which compares with the 816 days for OSIRIS-REx's rendezvous with Bennu (launched on September 8, 2016 and rendezvous on December 3, 2018)~\cite{wibben2020osiris}. 
As shown in the last plot of \fig{fig:state}, the angular displacement between the solar sail and NEO increases to produce a phase difference of more than one orbit.



The optimal solar sail cone angle is shown in \fig{fig:input}, where it is noted that the magnitude of the cone angle never reaches the \SI{90}{\degree} bounds. 
The solar sail is constantly maneuvering with respect to the direction of solar sail radiation. 
The optimal input for this nonlinear system does not follow the minimum-time, ``bang-bang'' structure found for systems affine in the control variable \cite{Bryson}.

\section{Communication}
\label{sec:communications}
Free space optical communication is a mature field whose utility for space-based communication is similarly well-established \cite{chan2006FSOComms}.
This work considers an ultra long-range satellite-to-satellite communication during the course of the mission, as well as a terrestrial link from geostationary orbit.

Suppose we have a commercial edge emitting 10\si{\watt} 850\si{\nano\meter} laser with lossless optics providing a 1\si{\centi\meter} diameter aperture.
With a lossless receiver with a 1\si{\centi\meter} diameter aperture stationed \SI{1e9}{\meter} from the receiver with no pointing error, the maximum optical power that can reach the receiver (Eq. \ref{eq:powerRX}) is 85.4\si{\femto\watt}, or \SI{3.7E5}{photons/\second}.
\begin{equation}\label{eq:powerRX}
    P_\text{R} = P_\text{T}\frac{A_\text{T}A_\text{R}}{\lambda^2R^2}\eta 
\end{equation}
With a 10\si{\nano\meter} optical filter centered about 850\si{\nano\meter}, we find (Table \ref{tab:photonCount}) that for a direct detection scheme, steps must be taken to avoid pointing the receiver directly at the sun or at highly reflective objects.

\begin{table}[t]
    \centering
    \begin{tabular}{c c}
        \toprule
        Source & Photons/Second \\
        \midrule
        Signal HIGH & \SI{3.7E5}{}\\
        Magnitude 5 A0Va & \SI{3.2E2}{}\\
        Direct Sunlight & \SI{3.4E14}{}\\
        Perfect Sunlight Reflector & \SI{7.7E6}{}\\
        \bottomrule
    \end{tabular}
    \caption{Incident photons for a receiver with a 1\si{\centi\meter} diameter aperture accompanied by a 10\si{\nano\meter} filter centered about 850\si{\nano\meter}. All values scale linearly with receiver aperture area.}
    \label{tab:photonCount}
\end{table}

\begin{align}
    \begin{split}\label{eq:ber}
        \text{BER} &= \left[\frac{1}{2}\sum_{i=n_T}^\infty p_Y^0(i)\right] + \left[\frac{1}{2}\sum_{i=0}^{n_T}p_Y^1(i)\right]\\
            &\approx \left[\frac{1}{2}\sum_{i=n_T+1}^\infty \mathcal{N}(i|\mu_0, \sigma_0^2)\right] \\
            &+ \left[\frac{1}{2}\sum_{i=0}^{n_T}\mathcal{N}(i|\mu_1, \sigma_1^2)\right]
    \end{split}
\end{align}
Care must be taken to avoid pointing at the sun, and strong reflectors can similarly complicate communication.
Per Equation \ref{eq:ber} and the analysis in \cite{sarbazi2020spadBER}, pointing this system at a magnitude 5 A0Va star while avoiding direct and reflected sunlight, a 16-SPAD array with PDE approaching 0.6 can theoretically achieve 50kbps downlink with a BER of $10^{-2}$ for 5\si{\watt} average power consumed by the transmitting laser.
Increasing the receiver aperture area by $10\times$ with otherwise identical hardware can attain up to 100kbps downlink---corresponding to roughly 37 received signal photons per bit---with a BER approaching $10^{-3}$ with a transmitter pointing accuracy of roughly $\pm45\si{\micro\radian}$, even with SPAD dead time on the order of singles of microseconds.
Alternative modulation schemes and coherent detection can enable greater communication rates with still lower power, though adjustments from the classical analog regime to photon counting are necessary \cite{chitnis2014spads}.


From GEO with atmospheric transmission $\eta\approx0.5$ talking to an Earth-based receiver with a  10\si{\centi\meter} aperture, the transmitter beam can have significantly greater divergence.
A minuscule 500\si{\micro\meter} transmit aperture can achieve data rates well in excess of 100kbps with a pointing accuracy on the order of 1\si{\milli\radian}.
Per Equation \ref{eq:powerRX}, increasing the transmitter or receiver aperture permits inversely scaled transmitter power for the same performance.
In other words, a commercially available 10\si{\watt} NIR source, paired with with modest transmit and receive apertures on the order of millimeters and tens of centimeters respectively can achieve data rates above 100kbps with a bit error rate of $10^{-3}$ with a photon counting-based receiver.
Prototypes of systems with similar size scale and pointing accuracy have previously been demonstrated~\cite{last2003toward,leibowitz2005256}.

\section{Computation and Storage}
\label{sec:computation}
Cell phones and the internet of things have led to a wide range of choices in energy efficient, physically small computation platforms \cite{samson2014small}, \cite{samson2011implementation}.
The candidate platform should have the computational capabilities to store and process high resolution images.  Other storage and processing requirements, such as guidance navigation and control, and communication processing, place significantly lower demands on processing and storage.
Even the image processing task has relatively low requirements, with allowable processing times on the order of minutes or perhaps even hours.

Outperforming these requirements is available in a variety of single-board computers, of which the VoCore2 PC~\cite{vocore2}is a representative. This \$50 computer runs Linux on a \SI{580}{MHz} MIPS 24K processor with 128 MB of RAM and 16 MB of flash. The entire computer weighs \SI{2.5}{\gram}, including connectors for camera and power, empty board space, and a microSD card slot. Adding a 512 GB card to store images and other data increases the system weight by roughly 0.25 grams. The processor burns roughly 1\si{\watt} peak, and has many power control modes and has a temperature operating range from 0\si{\celsius} to 85\si{\celsius}.


Similar computational capability in a carefully designed board with commercial off-the-shelf components could weigh roughly a gram, and operate from $-40$\si{\celsius} to 85\si{\celsius}. 
Custom designed silicon would drop the mass to well under a gram and increase the allowed temperature range while allowing radiation hardening by design.



\section{Radiation Effects}
\label{sec:radiation}

Ionizing radiation in interplanetary space is a challenge for electronics.  
The BLISS mass budget does not allow for sufficient shielding to have a significant impact on the dose hitting the electronics. 
Still, carefully-designed electronics should be able to survive a multi-year mission with sufficient probability that a swarm of spacecraft will achieve useful mission results.  
Multiple measurements put the normal radiation dose at 0.1 to 0.2 Gray per annum~\cite{reitz2012radiation, dachev2011overview,mrigakshi2013galactic}. 
Modern CMOS appears to be less susceptible to total ionizing dose than older technologies, perhaps due to gate oxides becoming so thin that trapped charges quickly tunnel away.  
Whatever the reason, it has been demonstrated that 14nm FINFETs can survive at high dosage levels. Hughes {\em et al.} ~\cite{hughes2015total} report leakage and threshold voltage shifts in several 14nm FET technologies after exposure to various levels of total ionizing doses. Their conclusions are that the electronics can survive roughly 100krad (1kGy) without serious effect, and up to 1 Mrad (10 kGy) with careful circuit design. 

Surviving single-event upsets will require redundancy in both hardware and software through either additional COTS components or custom circuitry.

\section{Discussion and Technology Trends}
\label{sec:limits}

The capabilities showcased here highlight the potential for swarms of low-cost spacecraft in the 10s of grams weight range with a relatively small form factor for observation, and possibly sample return, available with COTS. The time to escape from near-Earth orbit was approximately 120 days, swinging out to 2 AU in under a year, completing the interplanetary travel to intercept Bennu at approximately 1.3 AU in 640 days, orbiting for 346 days to collect images (similar orbit time as Osiris-REx), and returning to earth-orbit in similar time to intercept brings round-trip mission time to completion in just over 5.1 years (1,866 days) which is comparable to most small scale short missions but much faster than a similar Osiris-REx mission to an NEA which took just over 7 years (2,572 days). Although the Osiris-REx mission to Bennu included a significant time to select a site for material retrieval and subsequent material gathering it is still a valuable benchmark for a case study such as this, especially when considering overall flight time.

For a spacecraft that returns to earth orbit to communicate, the ultimate limits to spacecraft size are perhaps an order of magnitude smaller than what is proposed here. All of the major systems (camera, computation and storage, communication, power) can be engineered for a mass of roughly \SI{0.1}{\gram} each. For a spacecraft of total mass \SI{1}{\gram} a sail of only $\SI{0.1}{\meter^2}$ weighing only \SI{0.1}{\gram} will suffice.
This presents an opportunity in the field to gain an agile tool for observation capable of a broad scope of missions with a relatively quick turnaround. Bringing to the forefront the possibility of customized missions for each interested scientist, company, or otherwise interested party at a fraction of the current cost.
Direct spectrography and microanalysis have long been the metric for understanding material composition. To that end, this work also proposes that the BLISS spacecraft can retrieve cometary samples for analysis on Earth. Comets in the inner solar system such as 107P Wilson-Harrington are a prime target for missions and the trajectory will be similar to that of the NEA rendezvous described earlier in this work.





\section{Conclusion}
\label{sec:conclusion}

This paper presents a vision for the Berkeley Low-cost Interplanetary Solar Sail spacecraft using micro-scale actuators and small solar sails to explore NEOs.
The initial actuation, control, communication, and computation tools required to realize the BLISS project rely on privatization of spaceflight and the consumerization of micro-technologies.
The tools presented in this paper represent the initially available mechanisms, and with specific research the parameters for the spacecraft can be improved substantially, which would move the progress of the project towards first launch and allow exploration of more ambitious profiles for future missions.
In providing a large base of results demonstrating the feasibility of such technologies, we hope that is opens further investigation into novel spacecraft and mission profiles.

\section*{Acknowledgements}
The authors would like to thank the Berkeley Sensor and Actuator Center (BSAC) for its continued support. This project was accomplished in part with funding from the National Science Foundation Graduate Research Fellowship Program (NSF-GRFP), the National GEM Fellowship, and the Ford Predoctoral Fellowship.
Also, thanks to Raymond Chong and Beau Kuhn for participating in many useful conversations.

\bibliographystyle{unsrt}  
\bibliography{references}

\section*{Author Biographies}

\begin{wrapfigure}{l}{0.35\linewidth}
    \includegraphics[width=\linewidth]{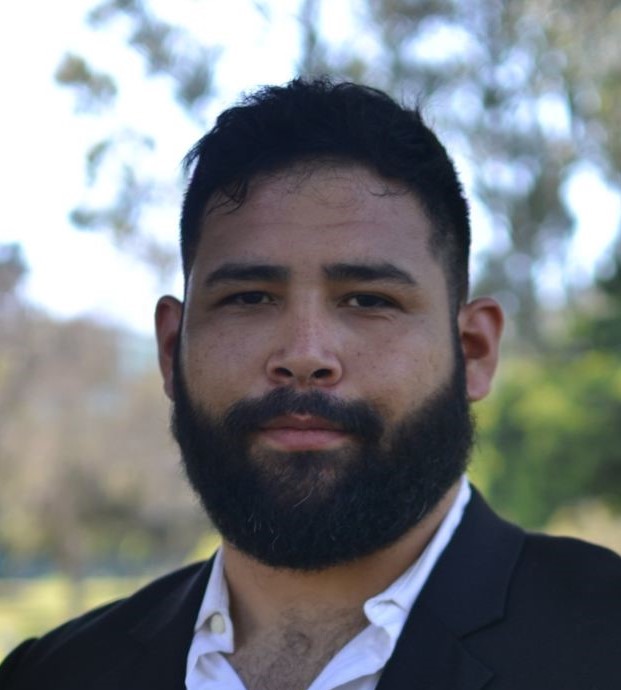}
\end{wrapfigure}
\textbf{Alexander N. Alvara} is a PhD student in the Department of Mechanical Engineering at University of California, Berkeley co-advised by Professor Liwei Lin and Professor Kristofer Pister with the Berkeley Actuators and Sensor Center. He received his three concurrent BS degrees in Mechanical Engineering, Aerospace Engineering, as well as Materials Science and Engineering at University of California, Irvine in 2017. Alexander's research is focused on micro-/nano-systems and materials with applications in extreme environments. 

\begin{wrapfigure}{l}{0.35\linewidth}
    \includegraphics[width=\linewidth]{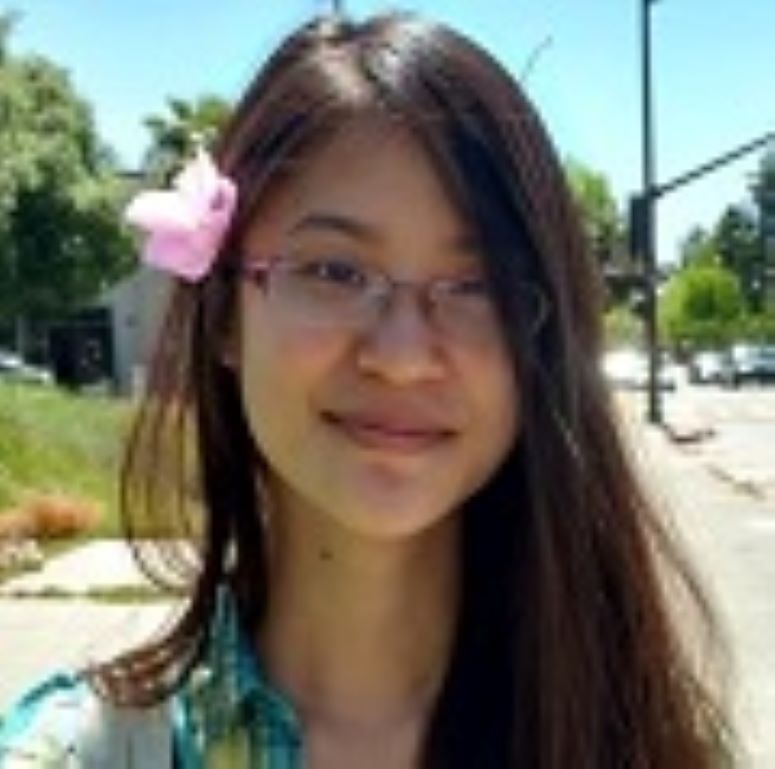}
\end{wrapfigure}
\textbf{Lydia Lee} is a PhD student in the Department of Electrical Engineering and Computer Sciences at University of California, Berkeley, advised by Professor Kristofer Pister in the Berkeley Autonomous Microsystems Lab. She received her BS in Electrical Engineering and Computer Sciences from University of California, Berkeley in 2017. Her research focuses on automated design of sensor front ends for low power electronics and space applications.

\begin{wrapfigure}{l}{0.35\linewidth}
    \includegraphics[width=\linewidth]{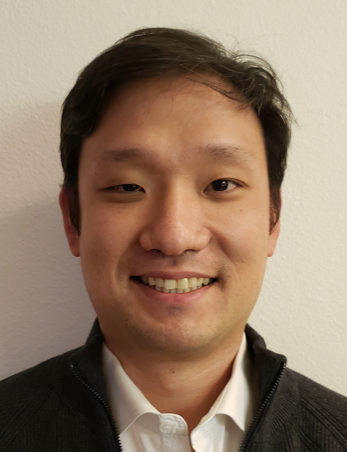}
\end{wrapfigure}
\textbf{Emmanuel Sin} recently finished his PhD in the Department of Mechanical Engineering at the University of California, Berkeley. 
He received his BS in Mechanical Engineering from Massachusetts Institute of Technology in 2007. His research interests are in aerospace guidance, navigation, and control. 

\begin{wrapfigure}{l}{0.35\linewidth}
    \includegraphics[width=\linewidth]{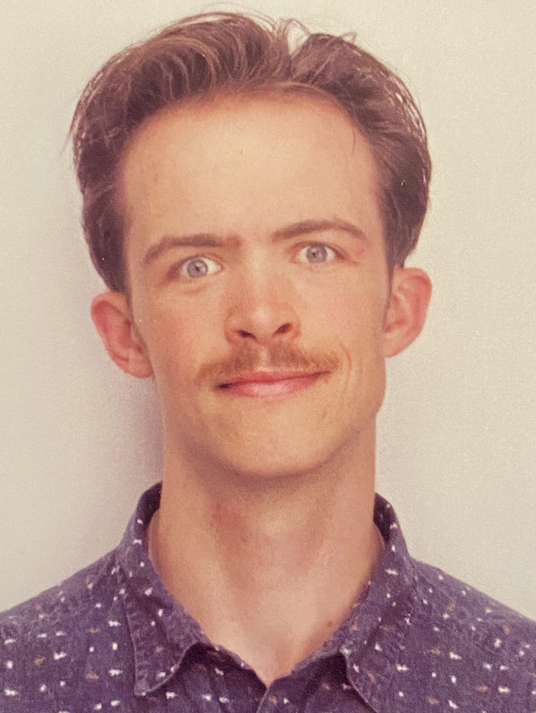}
\end{wrapfigure}
\textbf{Nathan Lambert} recently finished his PhD at the University of California, Berkeley. 
He was a member Department of Electrical Engineering and Computer Sciences, advised by Professor Kristofer Pister in the Berkeley Autonomous Microsystems Lab. 
His work explores many topics on model learning and decision making with data-driven and analytical methods. 
He received his BS in Electrical and Computer Engineering from Cornell University in 2017.


\begin{wrapfigure}{l}{0.35\linewidth}
    \includegraphics[width=\linewidth]{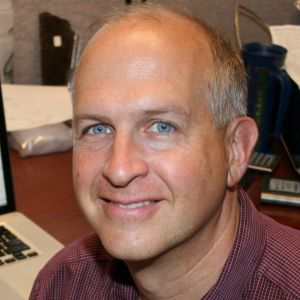}
\end{wrapfigure}
\textbf{Andrew Westphal} is a Research Physicist and Senior Space Fellow at the Space Sciences Laboratory at U. C. Berkeley. He got his PhD at Berkeley in 1992 in high-energy astrophysics, and now works on the interface between planetary science and astrophysics. The Westphal group uses some of the most sophisticated x-ray, electron-beam and ion-beam instruments on the planet to study ultra-primitive extraterrestrial materials that contain clues about the earliest history of the solar system. 

\begin{wrapfigure}{l}{0.35\linewidth}
    \includegraphics[width=\linewidth]{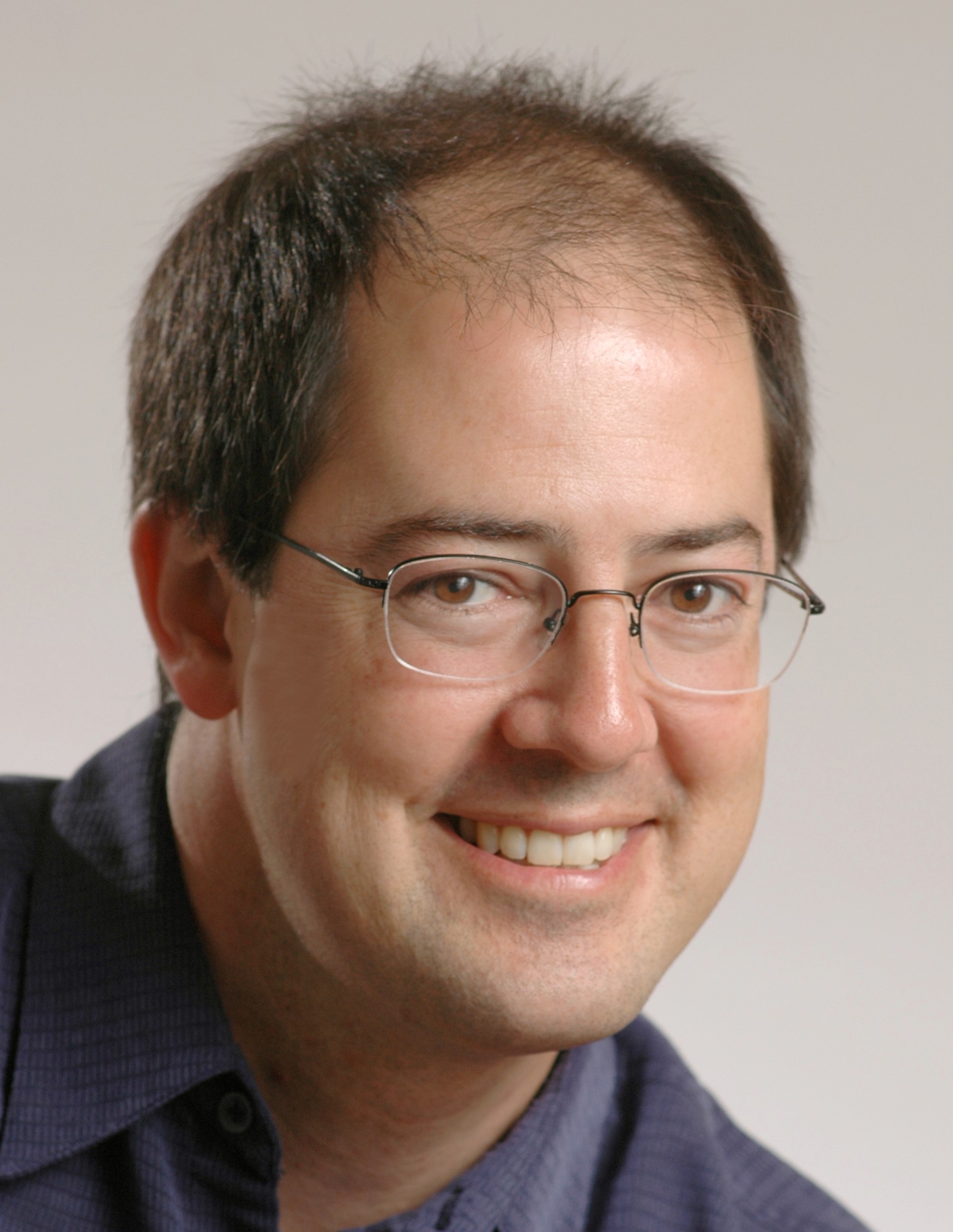}
\end{wrapfigure}
\textbf{Kristofer S. J. Pister} is a professor of Electrical Engineering and Computer Sciences at University of California, Berkeley and the founder and CTO of Dust Networks. He received a BA in Applied Physics from University of California, San Diego, 1986, and an M.S. and PhD in EECS from University of California, Berkeley in 1989 and 1992. Prior to joining the faculty of EECS in 1996, he taught in the Electrical Engineering Department, University of California, Los Angeles.


\end{document}